\documentclass[10pt]{article} 
\usepackage[utf8]{inputenc} 

\usepackage{geometry} 
\usepackage{graphicx} 
\usepackage[parfill]{parskip} 
\usepackage{booktabs} 
\usepackage{array} 
\usepackage{paralist} 
\usepackage{verbatim} 
\usepackage{subfig} 
\usepackage{authblk}

\usepackage{fancyhdr} 
\usepackage{sectsty}
\allsectionsfont{\sffamily\mdseries\upshape} 

\usepackage[nottoc,notlof,notlot]{tocbibind} 
\usepackage[titles,subfigure]{tocloft} 

\newcommand{\be}{\begin{equation}}
\newcommand{\ee}{\end{equation}}
\newcommand{\bea}{\begin{eqnarray}}
\newcommand{\eea}{\end{eqnarray}}

\title{The WIMP-Induced Gamma Ray Spectrum of \\ Active Galactic Nuclei}
\author[1]{M.A. G\'omez\thanks{miguel.gomezramirez@mavs.uta.edu}}
\author[1]{C.B. Jackson \thanks{chris@uta.edu}}
\author[2]{G. Shaughnessy \thanks{gshau@hep.wisc.edu}}
\affil[1]{Department of Physics, University of Texas at Arlington, Arlington, TX 76019, USA}
\affil[2]{Department of Physics, University of Wisconsin, Madison, WI 53706, USA}

\date{} 

\begin{document}

\maketitle

\begin{abstract}
As direct and indirect dark matter detection experiments continue to place stringent constraints on WIMP masses and couplings, it becomes imperative to expand the scope of the search for particle dark matter by looking in new and exotic places.  One such place may be the core of active galactic nuclei where the density of dark matter is expected to be extremely high.  Recently, several groups have explored the possibility of observing signals of dark matter from its interactions with the high-energy jets emanating from these galaxies.  In this work, we build upon these analyses by including the other components of the WIMP-induced gamma ray spectrum of active galactic nuclei; namely, ({\it i}) the continuum from WIMP annihilation into light standard model states which subsequently radiate and/or decay into photons and ({\it ii}) the direct (loop-induced) decay into photons.  We work in the context of models of universal extra dimensions (in particular, a model with two extra dimensions) and compute all three components of the gamma ray spectrum and compare with current data.  We find that the model with two extra dimensions exhibits several interesting features which may be observable with the Fermi gamma ray telescope.  We also show that, in conjunction with other measurements, the gamma ray spectrum from AGN can be an invaluable tool for restricting WIMP parameter space.
\end{abstract}

\newpage

\section{Introduction}
\label{sec:intro}

The search for dark matter has never experienced such exciting times.  Both direct and indirect detection experiments are starting to probe (and constrain) what many believe to be the preferred region of parameter space for particle dark matter (namely, electroweak-size couplings and masses).  There have also been several anomalies in the data from experiments on both sides that are consistent with signatures expected from particle dark matter, but may ultimately be explained away as either astrophysical in origin or as experimental errors.  

While these searches continue (and the debates over possible ``signals'' rage), it is important to make sure we are uncovering every stone in the search for dark matter by considering more exotic scenarios and/or looking in different locations in the universe.  Recently, several groups have investigated the possibility of detecting dark matter around active galactic nuclei (AGN) using the high-energy jets which emanate from these objects \cite{Bloom:1997vm,Profumo:2010,Tait:2012,Chang:2012sk,Profumo:2013jeb}.  AGN are thought to possess the densest distributions of dark matter in the universe which makes the probability of interactions between the jet particles (thought to be either electrons or protons) and dark matter particles in the halo surrounding the AGN core non-negligible.  The basic idea is that collisions of jet particles with WIMPs causes an ``up-scattering'' into a heavy, charged non-standard model particle.  This particle quickly decays back to the original jet and WIMP particles, but, in the process, a photon can be emitted from the initial, final or intermediate state charged particles.  The flux of gamma rays produced in this way has been shown to possess several interesting spectral features (including a sharp cutoff which depends on the mass splitting between the WIMP and the up-scattered particle) that would discriminate it from any astrophysical backgrounds.

With the expected high densities of WIMPs around the cores of AGN, the probabilities for {\it annihilations} of WIMPs could also be enhanced.  These annihilations can give rise to gamma rays in two ways.  The first is annihilation into lighter standard model (SM) states which then radiate photons or hadronize/decay into them (e.g., through neutral pion decay).  The result is a continuum which, generally, has a kinematic cutoff at the WIMP mass, but is otherwise featureless.  The other possible contribution to the gamma ray spectrum from WIMP annihilations comes from the loop-level processes which directly produce $\gamma + X$ final states (where $X$ can be another photon, a massive vector gauge boson or a scalar depending on the spin of the WIMP \cite{Bertone:2009cb,Jackson:2009kg,Bertone:2010fn,Jackson:2013pjq,Jackson:2013tca}).  Since the WIMPs are essentially at rest, the result of these annihilations are mono-energetic photons produced with energy:
\begin{equation}
E_\gamma = M_{DM} \left( 1 - \frac{M_X^2}{4 M_{DM}^2} \right) \,,
\label{eq:Egamma}
\end{equation} 
where $M_{DM}$ is the mass of the WIMP and $M_X$ is the mass of the particle produced in association with the photon.  Naively, since these processes occur at loop-level, one would assume that they are suppressed compared to the tree-level continuum annihilations.  However, it has been demonstrated for several models that the line-emission can compete with (and sometimes even dominate over) the continuum.  The focus of most of these types of studies has been either the center of the Milky Way galaxy or its orbitting satellite (dwarf) galaxies.  The characteristics of continuum and line emission (and their relative sizes), however, is strictly determined by the particle physics involved (and not the astrophysics).  Thus, the gamma ray spectrum of other astrophysical objects should exhibit similar features.  In any case, observation of line emissions would provide a smoking gun for indirect dark matter detection since astrophysical processes are incapable of producing such features.

In this paper, we compute the complete WIMP-induced gamma ray flux for an AGN (including the AGN jet-WIMP interaction component along with the contributions from annihilations).  Since previous studies focussed solely on the gamma ray spectrum coming from the interaction between AGN jets and WIMPs, this marks the first time that the complete spectrum has been constructed.  We choose, as a particular example, to work in the context of the six-dimensional ``chiral square'' model \cite{Dobrescu:2004zi}.  This choice is only for illustrative purposes and is due, in part, because the chiral square model has been shown to produce several distinguishable line emisssions from annihilations (a generic situation dubbed ``the WIMP forest'') \cite{Bertone:2009cb}.

The rest of the paper is organized in the following manner.  First, in Section \ref{sec:model}, we give a brief overview of the 6-d chiral square model and discuss constraints on the mass of the WIMP candidate in this model from relic density considerations.  In Section \ref{sec:AGN-dm}, we compute the gamma ray flux from WIMP interactions with AGN jets.  This calculation involves several pieces (which can be categorized as coming from astrophysics or from particle physics) and we will discuss each piece in some detail.  Section \ref{sec:annihilations} contains a discussion on the gamma ray flux coming from WIMP annihilations around AGN.  In Section \ref{sec:scan-results}, we combine the various contributions to the AGN gamma ray flux and compare with current observations fro the Fermi gamma ray telescope.  In this section, we also perform a numerical fit of the model parameters using the gamma ray flux from an AGN (as observed by the Fermi Gamma Ray telescope) along with the observed relic density and the most recent constraints from direct detection experiments.  Section~\ref{sec:discussion} discusses the implications of the fits and illustrates the usefullness of AGN gamma ray spectra not only as a way to look for signals of dark matter, but as a tool to constrain its parameter space.  Finally, in Section \ref{sec:conclusion}, we conclude.

\section{The Model}
\label{sec:model}

The ``chiral square'' model is a model of universal extra dimensions (UEDs) in which all SM fields are allowed to propagate in two extra dimensions.  The exact details of the model can be found in Refs. \cite{Dobrescu:2004zi,Burdman:2005sr,Burdman:2006gy,Dobrescu:2007xf}.  Here, we only point out features that are of interest to our study.

The two extra-dimensional coordinates can be represented by a pair of points $(x^5, x^6)$ living in a square region with sides $L$.  The extra dimensions are orbifolded in such a way that adjacent sides of the square are identified with each other, 
\begin{equation}
(y, 0) \equiv (0, y) \,\,\, , \,\,\, (y, L) \equiv (L, y) \,.
\label{eq:square-sides}
\end{equation}
The folding leaves the two corners of the square which lie along the fold (at $(0,0)$ and $(L,L)$) invariant, and identifies the remaining two corners (at $(0,L)$ and $(L,0)$) as the same point.

The Kaluza-Klein (KK) modes of the SM fields are lablelled by a pair of integers $(j,k)$ which satisfy:
\begin{equation}
k \ge 0 \,\,\, , \,\,\, j \ge 1 - \delta_{k,0} \,.\label{eq:KKnumbers}
\end{equation}
The masses of the KK modes are generally given by:
\begin{equation}
M_{(j,k)}^2 = M_0^2 + \pi^2 \frac{j^2 + k^2}{L^2}\label{eq:KKmass}
\end{equation}
where $M_0$ is the mass of the ``zero-mode'' field which we identify with the SM field.  The KK modes of fermions are Dirac particles, while the gauge fields decompose into 4-d vectors $V^\mu$ and two 4-d scalars, $V^5$ and $V^6$.  One linear combination of these scalars is eaten, level-by-level, by the vector KK modes to provide their longitudinal degrees of freedom.  The other linear combination of $V^5$ and $V^6$ are physical gauge adjoint scalars which we denote as $V_H^{(j,k)}$.

As is the case in the much more studied 5-d UED model, the chiral square model contains a ``lightest Kaluza-Klein particle'' (or LKP) which is forbidden (at tree-level) from decaying due to remnants of extra-dimensional spacetime symmetries.  Thus, the LKP (which is a (1,0) excitation in the notation introduced above) provides an excellent candidate for cold dark matter.  The exact identity of the LKP depends on boundary terms, but we will follow the usual reasoning and assume that these terms are such that colored and charged KK modes are heavier than neutral modes.  Following these (and other) assumptions, the lightest $(1,0)$ mode is expected to be the scalar partner of the hypercharge gauge boson, $B_H^{(1,0)}$ (which we will abbreviate as $B_H$). In general, electroweak symmetry breaking mixes $B_H$ with its $SU(2)$ counterpart, $W_H^{(1,0)}$.  However, the mixing angle is typically small and the LKP (to a very good approximation) is pure $B_H$.  This means that the LKP coupling to fermions and the SM Higgs is controlled exclusively by the $U(1)$ gauge coupling $g_1$ and the hypercharge of the matter field.  Of particular interest to the work performed here is the coupling between the WIMP, an electrically-charged SM fermion and its first KK counterpart.  The effective Lagrangian for this interaction can be written as:
\begin{equation}
\Delta {\cal{L}} = g_1 \left[ \bar{\psi}_E  ( Y_L P_L + Y_R P_R ) \psi_e + \bar{\psi}_e  ( Y_L P_R + Y_R P_L ) \psi_E \right] B_H \, ,
\end{equation}
where $Y_L (Y_R)$ is the hypercharge quantum number for a left-handed (right-handed) SM particle, $\psi_e$ is the SM fermion field and $\psi_E$ is its KK counterpart.

Since the LKP is a real scalar, it has mass-supppressed annihilations into light fermions and, thus, predominantly annihilates into pairs of heavy particles such as massive SM gauge bosons and, if heavy enough, top quarks.  These annihilations are dominated by $s$-channel exchange of the SM Higgs boson and, as a consequence, its thermal relic density is very sensitive to the mass of the Higgs $M_H$ (as well as its own mass $M_{B}$) \cite{Dobrescu:2007ec}.  In particular, assuming the recent discovery at the LHC of a spin-0 boson with mass $\sim 125$ GeV is indeed the SM Higgs boson, the mass of $B_H$ is tightly constrained to lie in the range $190$ GeV $ \le M_{B} \le 215$ GeV.  Note that this is assuming that $B_H$ is solely responsible for the current density of DM.  It could be that the LKP is not a thermal relic or that there are other quasi-stable particles which contribute to the relic density.  In these cases, the mass of $B_H$ may not be so tightly constrained.  We will investigate the constraints on the WIMP mass (and other parameters) further in Section \ref{sec:scan-results} using a sophisticated Markov Chain Monte Carlo technique.

\section{Gamma Rays from the Interaction of AGN Jets with the WIMP Halo}
\label{sec:AGN-dm}

In this section, we will compute the necessary components in order to predict the flux of gamma rays originating from the interaction of AGN jet particles with halo WIMPs.  For the remainder of this paper, we will assume that the jet particles are exclusively electrons.  The gamma ray flux originating from these types of interactions can be expressed as an integral over the energy of the impinging jet particles ($E_e$) as:
\begin{eqnarray}
\frac{d \Phi_\gamma}{d E_\gamma} = \int \,\delta_{DM} \,\times \left( \frac{1}{d_{AGN}^2} \frac{d \Phi^{AGN}_e}{d E_e} \right) \times \left( \frac{1}{M_{B}} \frac{d^2 \sigma_{e + B_H \to \gamma + e + B_H}}{d\Omega dE_\gamma}\Bigg|_{\theta=\theta_0} \right) dE_e \,,
\label{eq:gammaraysflux}
\end{eqnarray}
where the first term in the integral is related to the density of dark matter near the AGN core, the second factor accounts for the dynamics of the jet particles and the third factor is the cross section for the interaction of the jet particles with the halo WIMPs.  Note that the first two factors (which depend both explicitly and implicitly on the characteristics of a particular AGN such as its distance from Earth, $d_{AGN}$) determine the number of particles of each kind that can interact and the third factor (which depends on the scattering angle $\theta_0$ between the jet axis and the line of sight) determines how frequently they interact at a given energy for the jet particle. In the subsequent sub-sections, we will consider each of the three above factors in detail.  The discussion of the first two factors will follow closely the discussion in Refs. \cite{Profumo:2010} and \cite{Tait:2012} and is only included here for completeness (and as a sanity check for the authors).  Those readers familiar with the results of Refs. \cite{Profumo:2010} and \cite{Tait:2012} can safely skip to Section \ref{subsec:AGN-WIMP-xn}.

\subsection{Dark Matter Density Profile}
\label{subsec:DM-profile}

The first factor in Eq. \ref{eq:gammaraysflux} is defined by the line of sight integral (as observed by the particles in the jet) of the dark matter density profile $\rho_{DM}(r)$:
\begin{equation}
\delta_{DM}\equiv\int_{r_{min}}^{r_{0}}\rho_{DM}\left(r\right)dr \,,
\label{eq:intlosdensitydm}
\end{equation}
where $r_{min}$ is the minimum distance of interest (i.e., the ``base'' of the jet) and $r_0$ is the distance at which the jet becomes incohesive and irrelevant.  Previous studies have found that, while results depend sensitively on the value of $r_{min}$, the actual value of $r_0$ plays little or no role (since dark matter density profiles typically fall off steeply with increasing radius).

Although little is known about the dark matter density profiles of active galaxies, there are several models on the market.  In this work, we focus on one particular model proposed by Gondolo and Silk \cite{Gondolo:1999ef}.  In the Gondolo-Silk model (which assumes collisionless dark matter), the central black hole is assumed to grow adiabatically by the accretion of surrounding gas and stars.  The result is a dark matter density profile with a very dense central spike.  Assuming an initial dark matter density distribution with a power-law profile ($\rho(r) \sim r^{-\gamma}$), the Gondolo-Silk profile takes the form:
\begin{equation}
\rho\left(r\right)=\frac{\rho'\left(r\right)\;\rho_{core}}{\rho'\left(r\right)+\rho_{core}} \,.
\label{eq:GS-profile}
\end{equation}
The density in the core is dominated by $\rho_{core}$ which depends sensitively on the mass of dark matter and the age of the central black hole ($t_{BH}$) as $\rho_{core}\simeq M_{DM}/\left(\langle\sigma\,v\rangle_{0}\,t_{BH}\right)$.  The other component of the Gondolo-Silk profile is given by:
\begin{equation}
\rho'\left(r\right)=\rho_{0}\left(\frac{R_{sp}}{r_{0}}\right)^{-\gamma}\left(1-\frac{4 R_{S}}{r}\right)^{3}\left(\frac{R_{sp}}{r}\right)^{\gamma_{sp}}
\end{equation}
where $\gamma_{sp}$ is the slope of the density profile of the central spike, $R_{sp}$ is the radius of the central spike and $R_{S}$ is the Schwarzschild radius of the black hole. The spike slope and radius are given respectively by:
\begin{equation}
\gamma_{sp} = \frac{9 - 2\gamma}{4 - \gamma} \,\,\,\,\, , \,\,\,\,\,
R_{sp}=\alpha_{\gamma}\,r_{0}\,\left(\frac{M_{BH}}{\rho_{0}\,r_{0}^{3}}\right)^{1/\left(3-\gamma\right)}
\label{eq:Rsp}
\end{equation}
and $\alpha_{\gamma}\propto\gamma^{4/9}$ \cite{Tait:2012}. Note that, because no stable orbit exists within the radius $r < 4 R_S$,  $\rho_{DM}\left(r<4\,R_{S}\right)=0$.  This condition sets a natural lower limit of integration $r_{min}=4\,R_{S}$ for Eq. \ref{eq:intlosdensitydm}. Finally, to fix the value of $\rho_{0}$, the normalization condition:
\begin{equation}
\int_{4\,R_{S}}^{10^5\,R_{S}}4\pi r^{2}\rho_{DM}\left(r\right)dr\leq\Delta M_{BH} \,,
\end{equation}
is used.  In other words, we require that the mass coming from dark matter be (at most) as large as the uncertainty in the mass of the central black hole ($\Delta M_{BH}$).  For the remainder of this paper, we will concentrate on the Centaurus A AGN.  The parameter choices we use are~\cite{Falcone:2010fk}:

\begin{center}
  \begin{tabular}{ l  r }
    \hline
    \hline
   $M_{BH}$  [ $M_\odot$ ] Black Hole mass  & $(5.5 \pm 3.0) \times 10^7$ \\
   $R_S$ [pc] Schwarzschild radius & $5 \times 10^{-6}$ \\
   $t_{BH}$ [yr] Age of Black Hole & $10^8 - 10^{10}$ \\
   $\alpha_\gamma$ & 0.1 \\
   $r_0$ [kpc] upper limit of integration & 15 \\
   $d_{AGN}$ & 3.7 Mpc \\
   $\theta_0$ & $68^\circ$ \\
    \hline
    \hline
  \end{tabular}
\end{center}

\begin{figure}[t]
\includegraphics[scale=0.26]{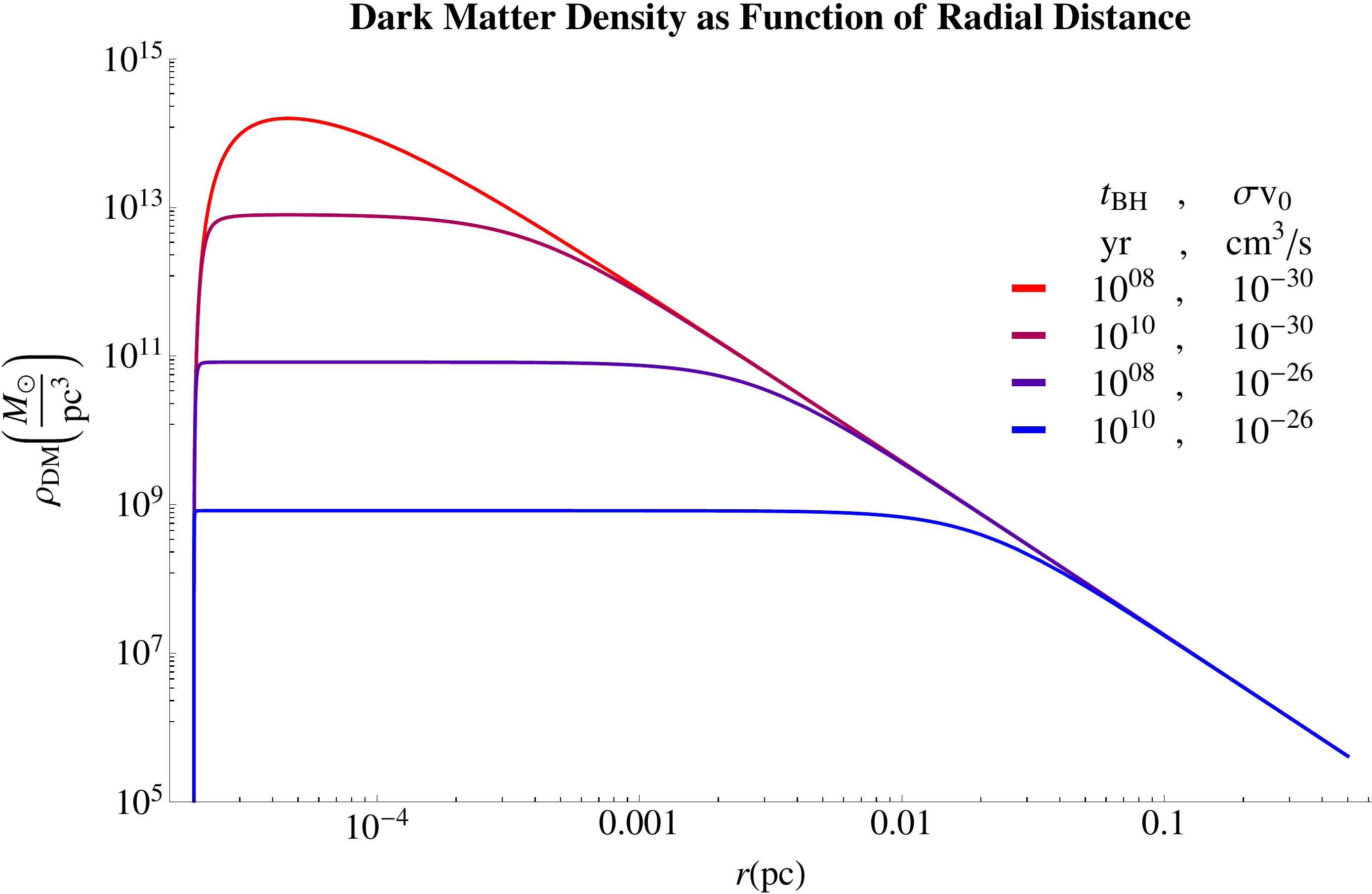}\hspace{0.25cm}
\includegraphics[scale=0.26]{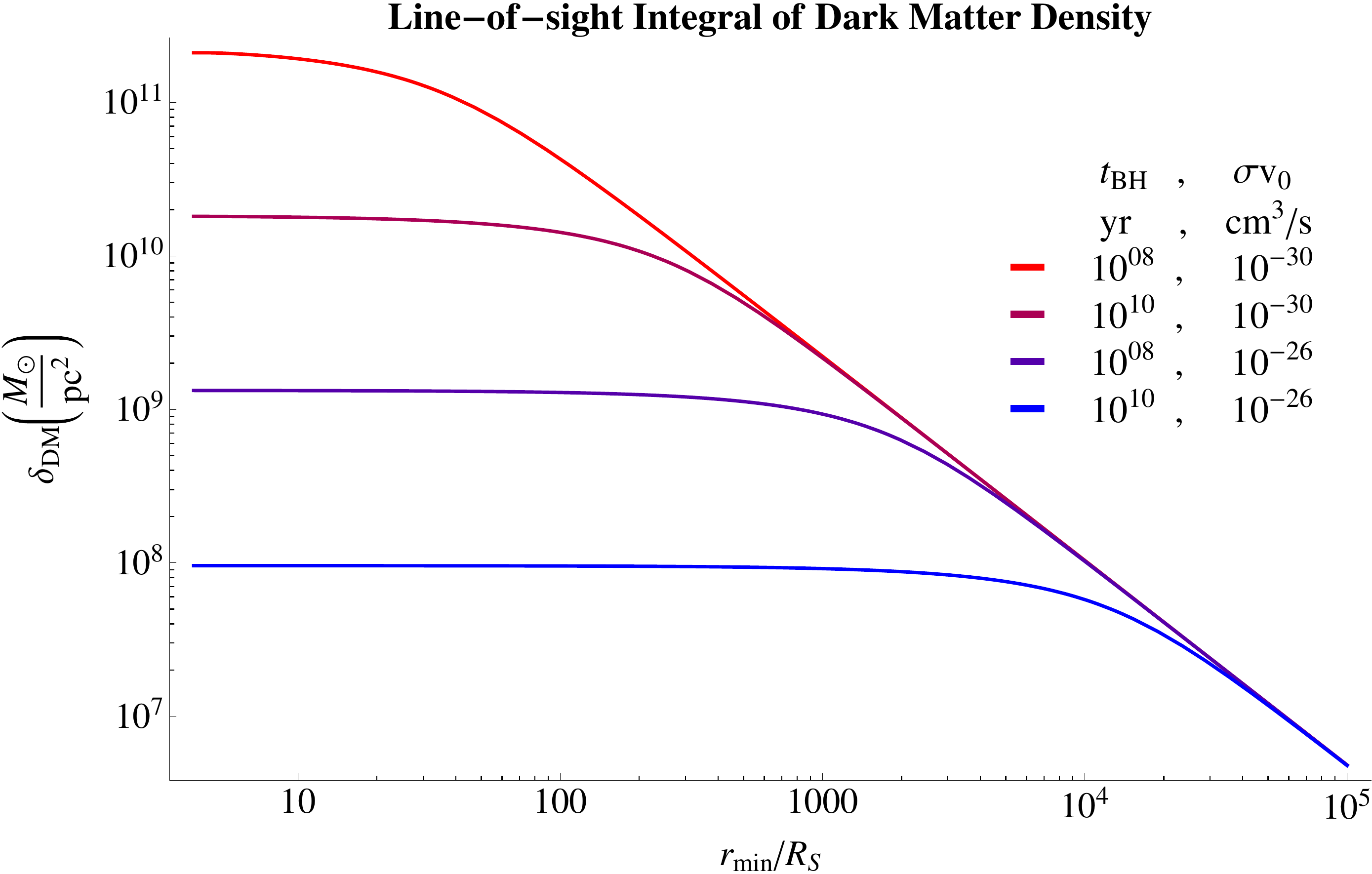}
\caption{Dark matter density profile for Centaurus A for different values of $\rho_{core}$ where $\gamma=1$, $M_{BH}=\left(5.5\pm3.0\right)\times10^{7}\,M_{\odot}$, $R_{S}=5\times10^6$ pc and $r_{0}=15\times10^3$ pc. Left dark matter density profile and right integral of line of sight of dark matter density profile.}\label{fg:dmdensity}
\end{figure}

The dark matter density profile for Centaurus A is shown on the left side of Figure \ref{fg:dmdensity} including four different scenarios depending on the value of the age of the black hole and the dark matter annihilation cross-section (ranging from $10^{-30}-10^{-26}$ cm$^3$/s) for a WIMP with a mass of 100 GeV.  On the right side of the same figure, we also show the line of sight integral ($\delta_{DM}$) for the same four scenarios.  As was first noted in \cite{Profumo:2010}, we see that a lower cross-section for pair annihilation allows a higher concentration of dark matter close to the spike.  This enhances the probability for an interaction between a jet particle and a halo WIMP.  We also see that younger black holes lead to a similar enhancement (due to there being less time for the black hole to deplete WIMPs living in the spike).  Finally, we see that the effective density that the jet probes varies very slowly with the lower limit of the integral (due to the spike in the density profile), but decreases rather quickly outside of the spike region.

\subsection{The AGN Jet Factor}
\label{subsec:AGN-jet}

Next, we discuss the second factor in Eq. (\ref{eq:gammaraysflux}) which accounts for the dynamics of the jet particles.  As pointed out in Refs.~\cite{Profumo:2010} and \cite{Tait:2012}, the exact geometry of the AGN jet is not very important for the purposes of our study.  In fact, the most important factor affecting the behavior of the jet for the interaction studied in this work is the energy of the particles in the jet. Using the blob geometry as defined in Ref.~\cite{herfbh} in which particles move isotropically and assuming that the gamma factor with respect to the black hole reference frame is not too large ($\Gamma_{B}\sim3$) a broken power law energy distribution is obtained from Fermi LAT observations \cite{Falcone:2010fk}:
\begin{equation}
\frac{d\Phi_{e}^{AGN}}{d\gamma^{\prime}}\left(\gamma^{\prime}\right)=\frac{1}{2}k_{e}\gamma^{\prime -s_{1}}\left[1+\left(\frac{\gamma^{\prime}}{\gamma^{\prime}_{br}}\right)^{\left(s_{2}-s_{1}\right)}\right]^{-1}\;\;\left(\gamma^{\prime}_{min}\leq\gamma^{\prime}\leq\gamma^{\prime}_{max}\right) \,,
\label{eq:kenorm}
\end{equation}
with $s_{1}=1.8$, $s_{2}=3.5$, $\gamma^{\prime}_{br}=4\times10^{5}$, $\gamma^{\prime}_{min}=8\times10^{2}$ , $\gamma^{\prime}_{max}=10^{8}$, and $\gamma^{\prime}=E^{\prime}/m_{e}$. The constant $k_{e}$ must be found from the kinetic power of the jet ($L_{e}$):
\begin{equation}
L_{e}=m_{e}\int_{-1}^{1}\int_{\gamma_{min}}^{\gamma_{max}}\frac{\gamma}{\Gamma_{B}\left(1-\beta_{B}\mu\right)}\frac{d\Phi_{e}^{AGN}}{d\gamma}\left(\gamma\Gamma_{B}\left(1-\beta_{B}\mu\right)\right)d\gamma d\mu \,,
\end{equation}
where the black hole reference frame is associated with unprimed quantities and the bulk reference frame with primed ones\footnote{$L_{e}$ is taken as the Eddington limit for the black hole mass.}.

The final form of the second factor in Equation \ref{eq:gammaraysflux} is then:
\begin{equation}
\frac{1}{d_{AGN}^{2}}\frac{d\Phi_{e}^{AGN}}{dE_{e}}=\frac{1}{d_{AGN}^{2}m_{e}}\int_{\mu_0}^{1}\frac{1}{\Gamma_{B}\left(1-\beta_{B}\mu\right)}\frac{d\Phi_{e}^{AGN}}{d\gamma}\left(\gamma\Gamma\left(1-\beta_{B}\mu\right)\right)d\mu
\label{eq:AGNjetfac}
\end{equation}
where $\mu_0$ parameterizes the jet collimation.  In the following, we adopt a value of $\mu_0 = 0.9$ to ensure a highly-collimated jet.

\subsection{The Cross Section}
\label{subsec:AGN-WIMP-xn}

\begin{figure}[t]
\includegraphics[scale=0.3]{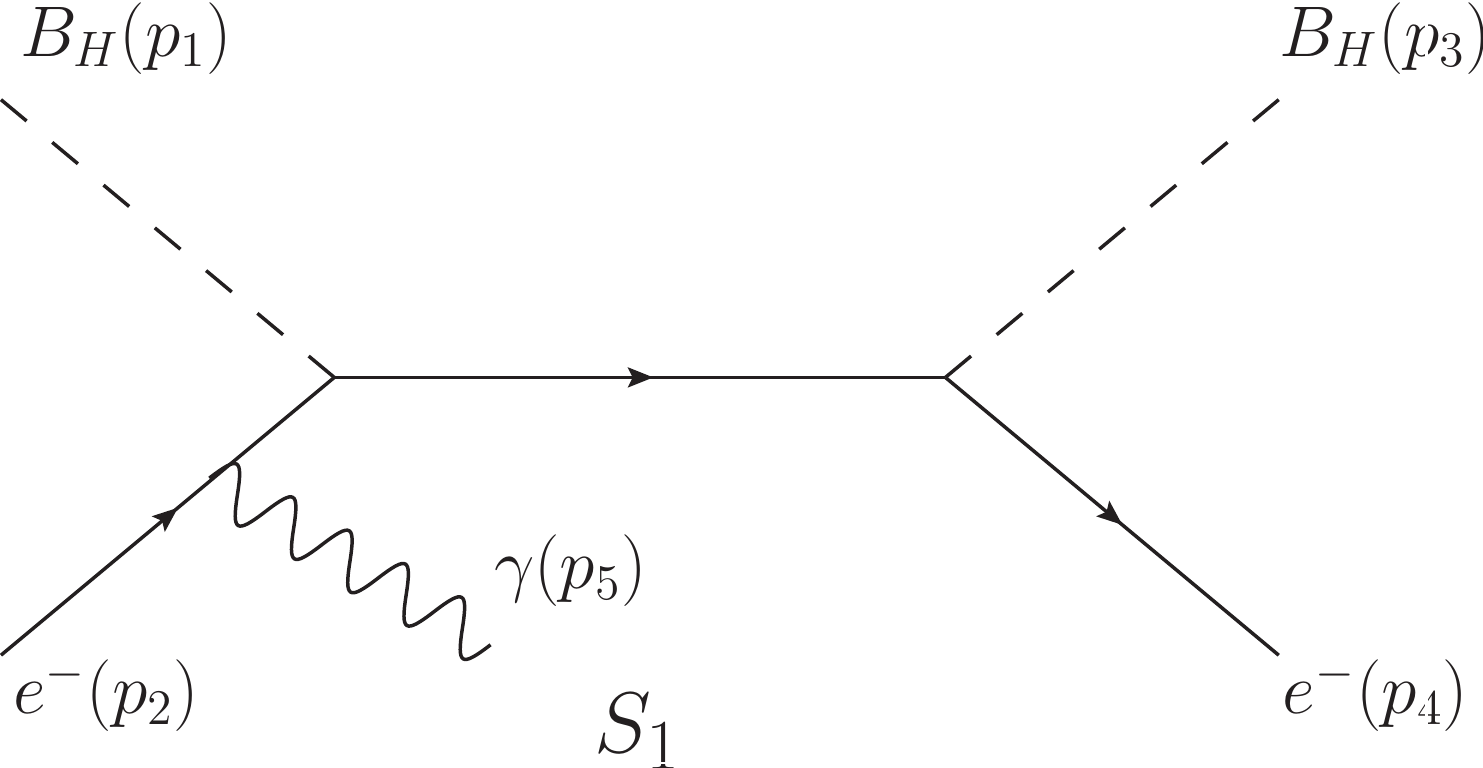}
\hspace{0.5cm}
\includegraphics[scale=0.3]{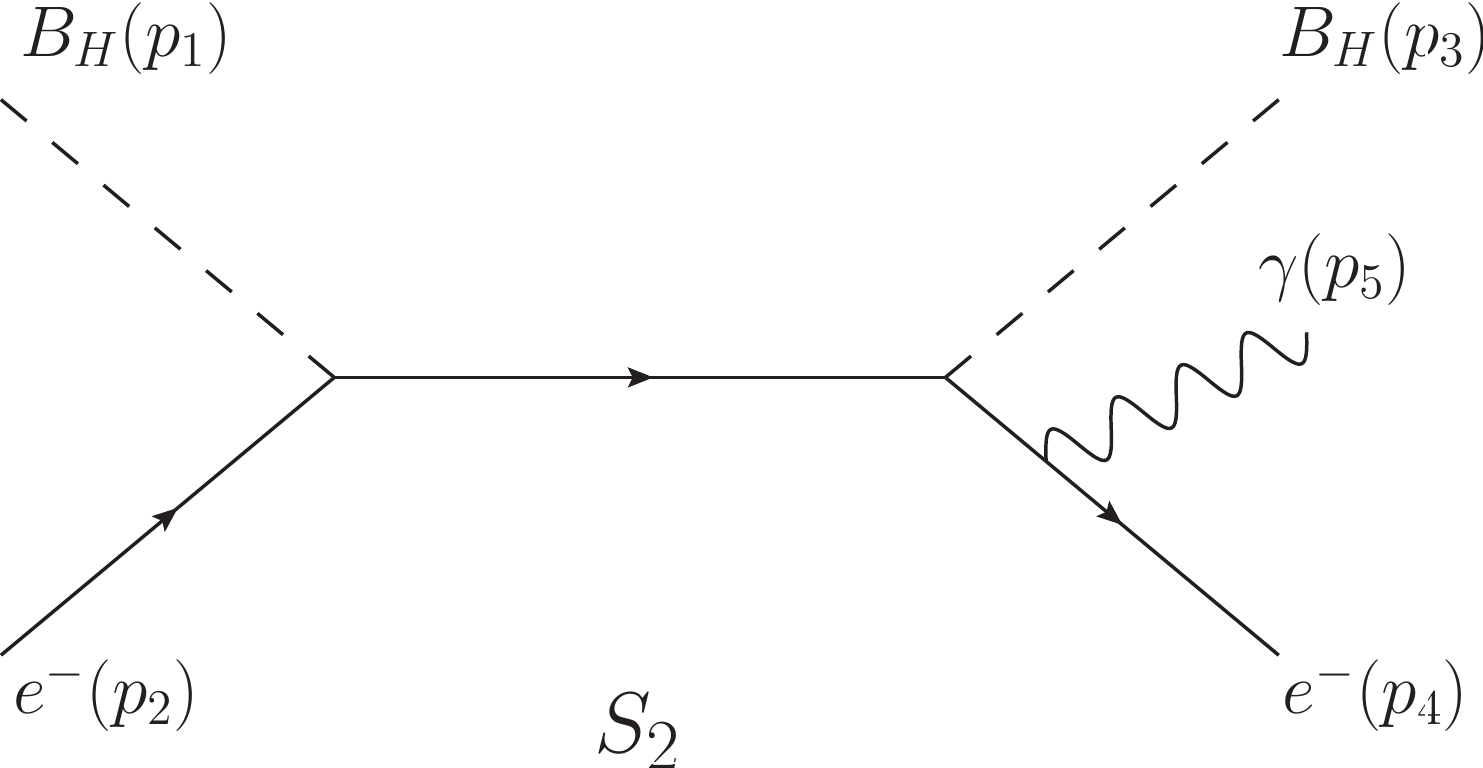}
\hspace{0.5cm}
\includegraphics[scale=0.3]{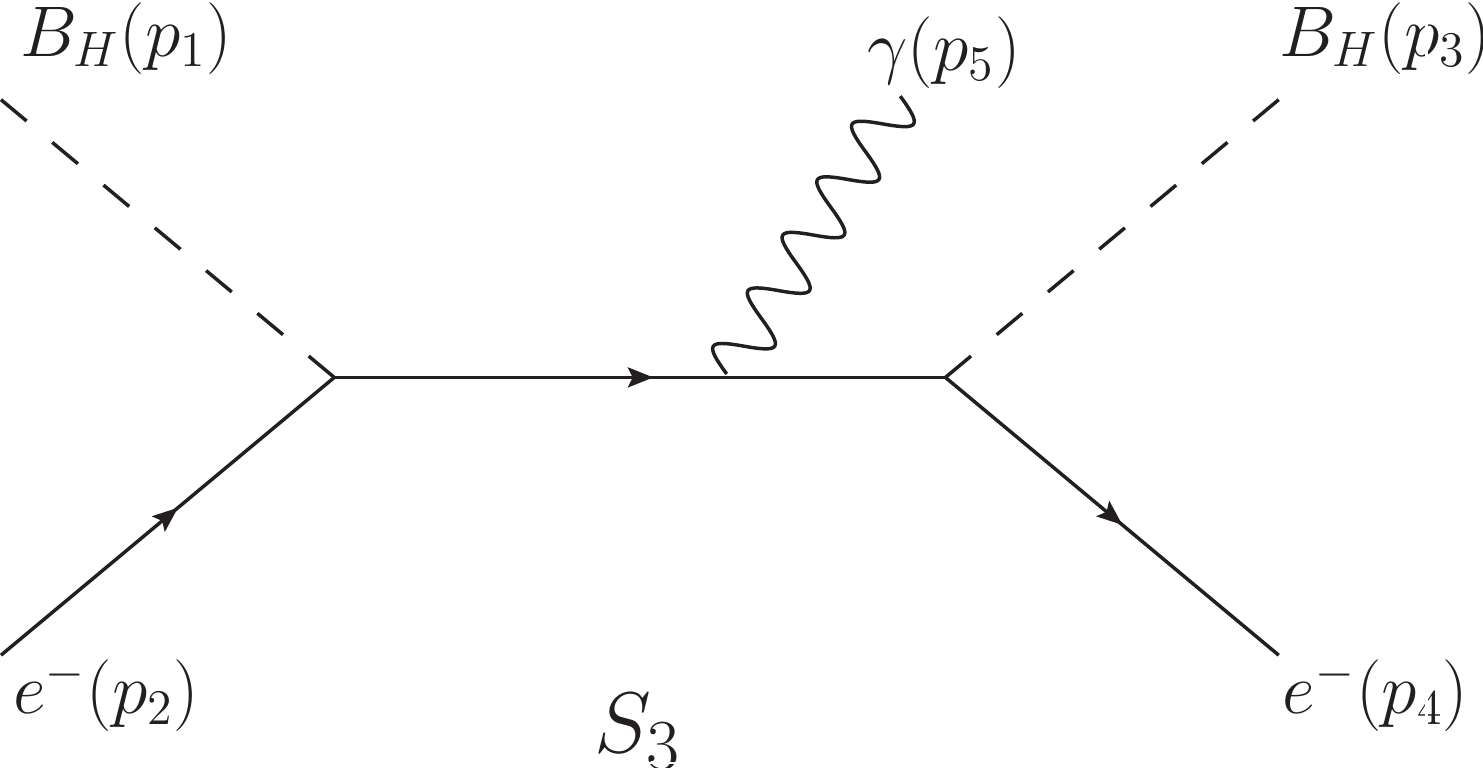} \\
\\
\\
\includegraphics[scale=0.3]{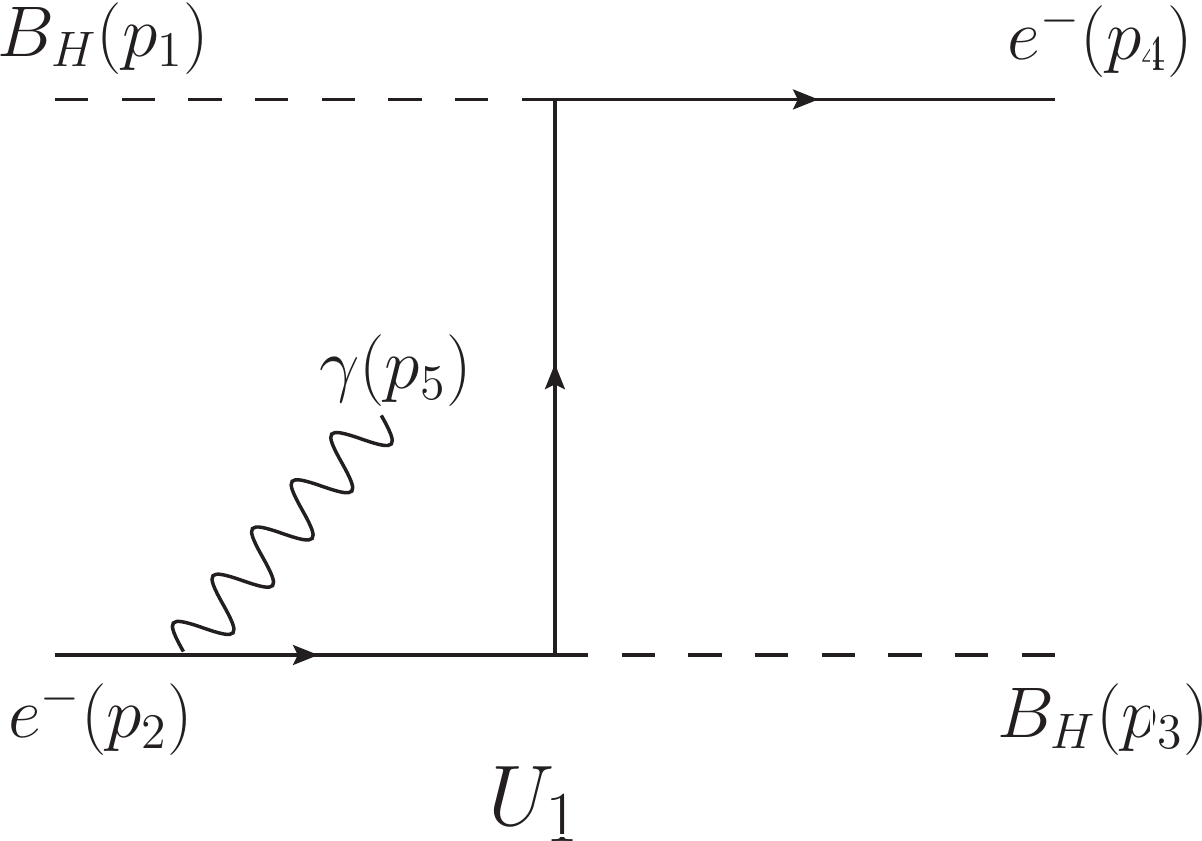}
\hspace{1.5cm}
\includegraphics[scale=0.3]{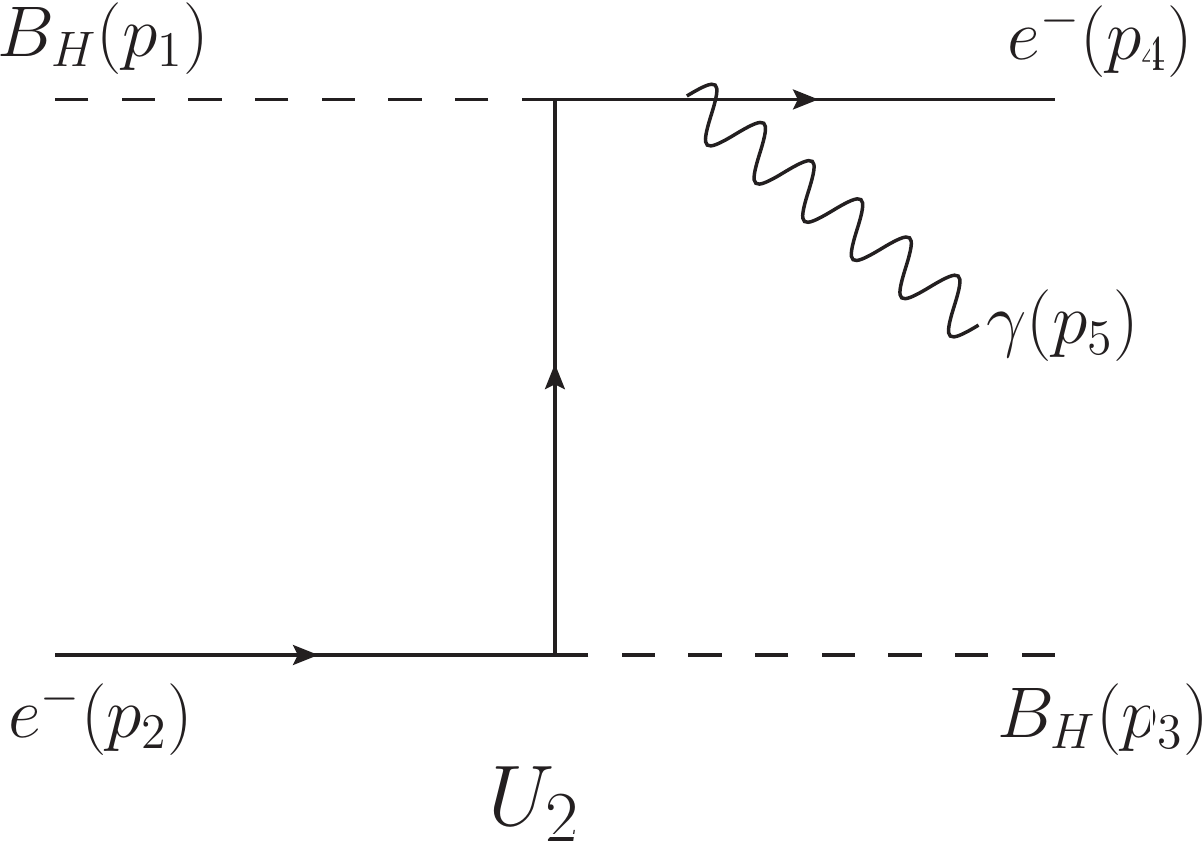}
\hspace{1.5cm}
\includegraphics[scale=0.3]{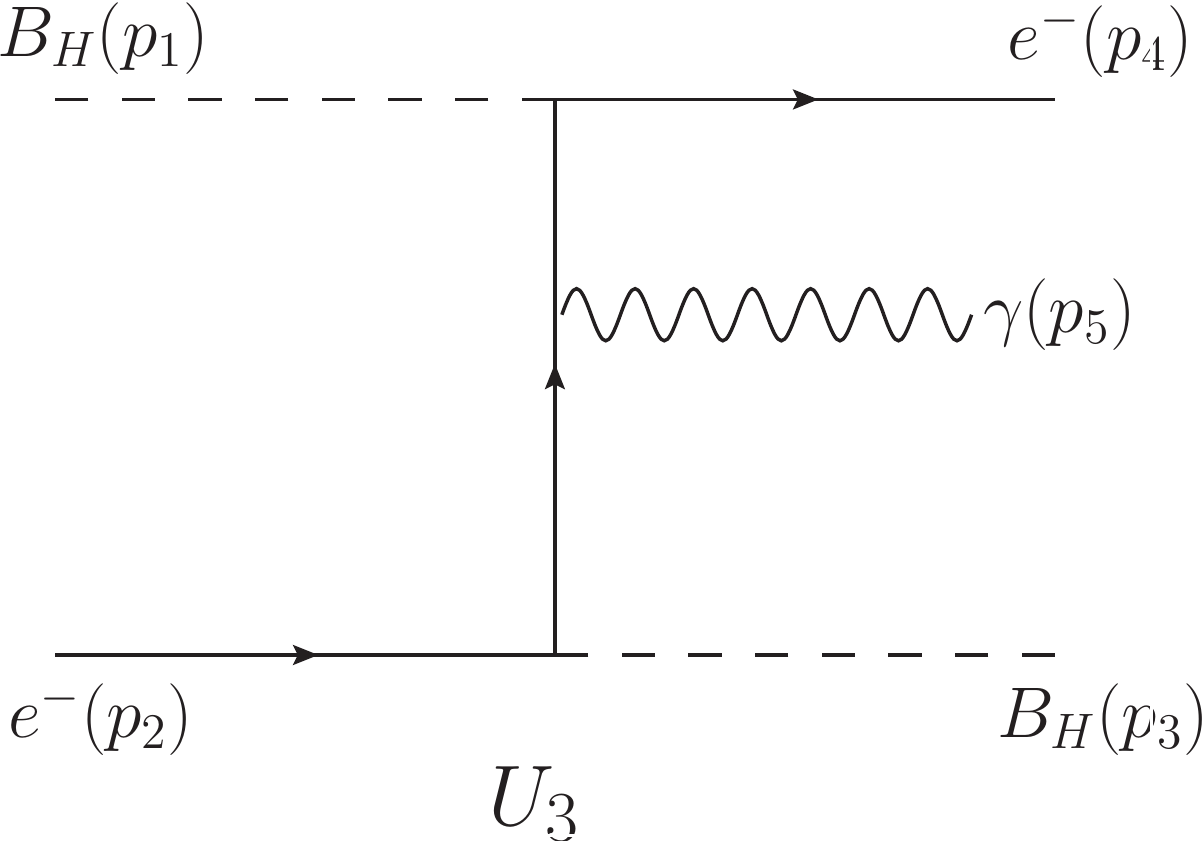}
\caption{Feynman diagrams for the process $B_H + e^- \to B_H + e^- + \gamma$ through $E$ exchange.}
\label{fg:feynman-diagrams}
\end{figure} 

Finally, we consider the third factor in Eq. (\ref{eq:gammaraysflux}) (i.e., the differential cross section for the process $B_H + e^- \to B_H + e^- + \gamma$).  The Feynman diagrams which contribute to $B_H + e^- \to B_H + e^- + \gamma$ are depicted in Fig.~\ref{fg:feynman-diagrams} where the virtual particle is taken to be $E^{(1,0)}$ (the first KK excitation of the electron which we will simply denote as $E$).  We write the total amplitude for this process as ${\cal M} = \sum_i {\cal M}_i^\mu \epsilon_\mu(p_5)$ where $\epsilon_\mu(p_5)$ is the polarization vector for the photon.  The amplitudes for the $s$-channel diagrams are given by:
\begin{eqnarray}
{\cal{M}}_{S_1} &=& \frac{e g_1^2}{\Sigma_{34} t_{25}} \bar{u}(p_4)  (Y_L P_R + Y_R P_L) (\not{p}_3 + \not{p}_4 + M_E) (Y_L P_L + Y_R P_R) (\not{p}_2 - \not{p}_5) \gamma^\mu u(p_2) \, , \\
\nonumber\\
{\cal{M}}_{S_2} &=& \frac{e g_1^2}{\Sigma_{12} s_{45}} \bar{u}(p_4) \gamma^\mu (\not{p}_4 + \not{p}_5 ) (Y_L P_R + Y_R P_L) (\not{p}_1 + \not{p}_2 + M_E) (Y_L P_L + Y_R P_R) u(p_2) \, , \\
\nonumber\\
{\cal{M}}_{S_3} &=& \frac{e g_1^2}{\Sigma_{12} \Sigma_{34}} \bar{u}(p_4) (Y_L P_R + Y_R P_L) (\not{p}_3 + \not{p}_4 + M_E) \gamma^\mu (\not{p}_1 + \not{p}_2 + M_E) (Y_L P_L + Y_R P_R) u(p_2) \,, \nonumber\\
\end{eqnarray}
where:
\begin{eqnarray}
\Sigma_{12} &=& s_{12} - M_E^2 - i \sqrt{s_{12}} \Gamma_E \, ,\\
\Sigma_{34} &=& s_{34} - M_E^2 - i \sqrt{s_{34}} \Gamma_E \, ,
\end{eqnarray}
and the kinematic invariants are defined by $s_{ij} = (p_i + p_j)^2$ and $t_{ij} = (p_i - p_j)^2$.  The energy-independent decay width $\Gamma_E$ for $E$ is computed to be:
\begin{equation}
\Gamma_E =  \frac{g_1^2}{32 \pi} (Y_L^2 + Y_R^2) \frac{(M_E^2 - M_B^2)^2}{M_E^3} \, .
\end{equation}
The $u$-channel amplitudes are:
\begin{eqnarray}
{\cal{M}}_{U_1} &=& \frac{e g_1^2}{t_{14} t_{25}} \bar{u}(p_4) (Y_L P_R + Y_R P_L) (\not{p}_4 - \not{p}_1 + M_E) (Y_L P_L + Y_R P_R) (\not{p}_2 - \not{p}_5) \gamma^\mu u(p_2) \,, \\
\nonumber\\
{\cal{M}}_{U_2} &=& \frac{e g_1^2}{t_{23} t_{45}} \bar{u}(p_4) \gamma^\mu (\not{p}_4 + \not{p}_5) (Y_L P_R + Y_R P_L) (\not{p}_2 - \not{p}_3 + M_E) (Y_L P_L + Y_R P_R) u(p_2) \,, \\
\nonumber\\
{\cal{M}}_{U_3} &=& \frac{e g_1^2}{t_{14} t_{23}} \bar{u}(p_4) (Y_L P_R + Y_R P_L) (\not{p}_4 - \not{p}_1 + M_E) \gamma^\mu (\not{p}_2 - \not{p}_3 + M_E) (Y_L P_L + Y_R P_R) u(p_2) \,. \nonumber \\
\end{eqnarray}

The differential cross section for $B_H(p_1) + e(p_2) \to B_H(p_3) + e(p_4) + \gamma^\mu(p_5)$ is given in general by:
\begin{equation}
d\sigma = (2\pi)^4 \delta^4(p_1 + p_2 - p_3 - p_4 - p_5) \frac{1}{4 E_1 E_2} \frac{d^3p_3}{(2\pi)^3 2 E_3}\frac{d^3p_4}{(2\pi)^3 2 E_4}\frac{d^3p_5}{(2\pi)^3 2 E_5} \overline{\sum} | {\cal{M}} |^2 \,.
\end{equation}
Using the $\delta^3(\vec{p_1} + \vec{p_2} - \vec{p_3} - \vec{p_4} - \vec{p_5})$ to integrate over $d^3p_3$ and the remaining $\delta(E_1 + E_2 - E_3 - E_4 - E_5)$ to integrate over $dE_4$, we arrive at a differential cross section of the form:
\begin{equation}
\frac{d\sigma}{dE_5 d\Omega_5} = \frac{1}{(2\pi)^5} \frac{E_5}{32 E_1 E_2} \int d\Omega_4 \frac{E_4}{E_3} \frac{1}{|1+J|} \sum | {\cal{M}} |^2 \,. \\
\label{eq:cross-sec}
\end{equation}
We choose to work in the initial WIMP rest frame and align our coordinate system such that the initial electron four-momentum lies along the $+\hat{z}$-axis and the azimuthal angle for the photon is zero ($\phi_5 = 0$).  In this frame, the relevant momenta are given by:
\begin{eqnarray}
p_1 &=& (M_B, \vec{0}) \\
p_2 &=& E_2 (1, 0, 0, 1) \\
p_4 &=& E_4 ( 1, \sin\theta_4 \cos\phi_4, \sin\theta_4 \sin\phi_4, \cos\theta_4) \\
p_5 &=& E_5 (1, \sin\theta_5, 0, \cos\theta_5)
\end{eqnarray}
and we use conservation of four-momentum to replace $p_3$ (i.e., $p_3 = p_1 + p_2 - p_4 - p_5$).  Note that, for Cen A, $\theta_5 = \theta_0 = 68^\circ$.  The energies for the final state WIMP ($E_3$) and electron ($E_4$) in this frame are given respectively by:
\begin{eqnarray}
E_3 &=& \biggl[M_B^2 + E_2^2 + E_4^2 + E_5^2 - 2 E_2 E_4 \cos\theta_4 - 2 E_2 E_5 \cos\theta_4 \nonumber\\
&&\,\,\,\,\,\,\,\,\,\,\,\,\,\,\,\, + \,\,\, 2 E_4 E_5 ( \sin\theta_5 \sin\theta_4 \cos\phi_4 + \cos\theta_5 \cos\theta_4)\biggr]^{1/2} \\
\nonumber\\
E_4 &=& \frac{M_B (E_2 - E_5) - E_2 E_5 (1-\cos\theta_5)}{ M_B + E_2 (1-\cos\theta_4) - E_5 (1 - \sin\theta_5 \sin\theta_4 \cos\phi_4 - \cos\theta_5 \cos\theta_4)}
\end{eqnarray}
while the Jacobian $J$ takes the form:
\begin{equation}
J = \frac{1}{E_3} \left[ E_4 - E_2 \cos\theta_4 + E_5 \left( \sin\theta_5 \sin\theta_4\cos\theta_4 + \cos\theta_5\cos\theta_4 \right) \right]\,.
\end{equation}

\begin{figure}[t]
\centering
\includegraphics[width=0.75\textwidth]{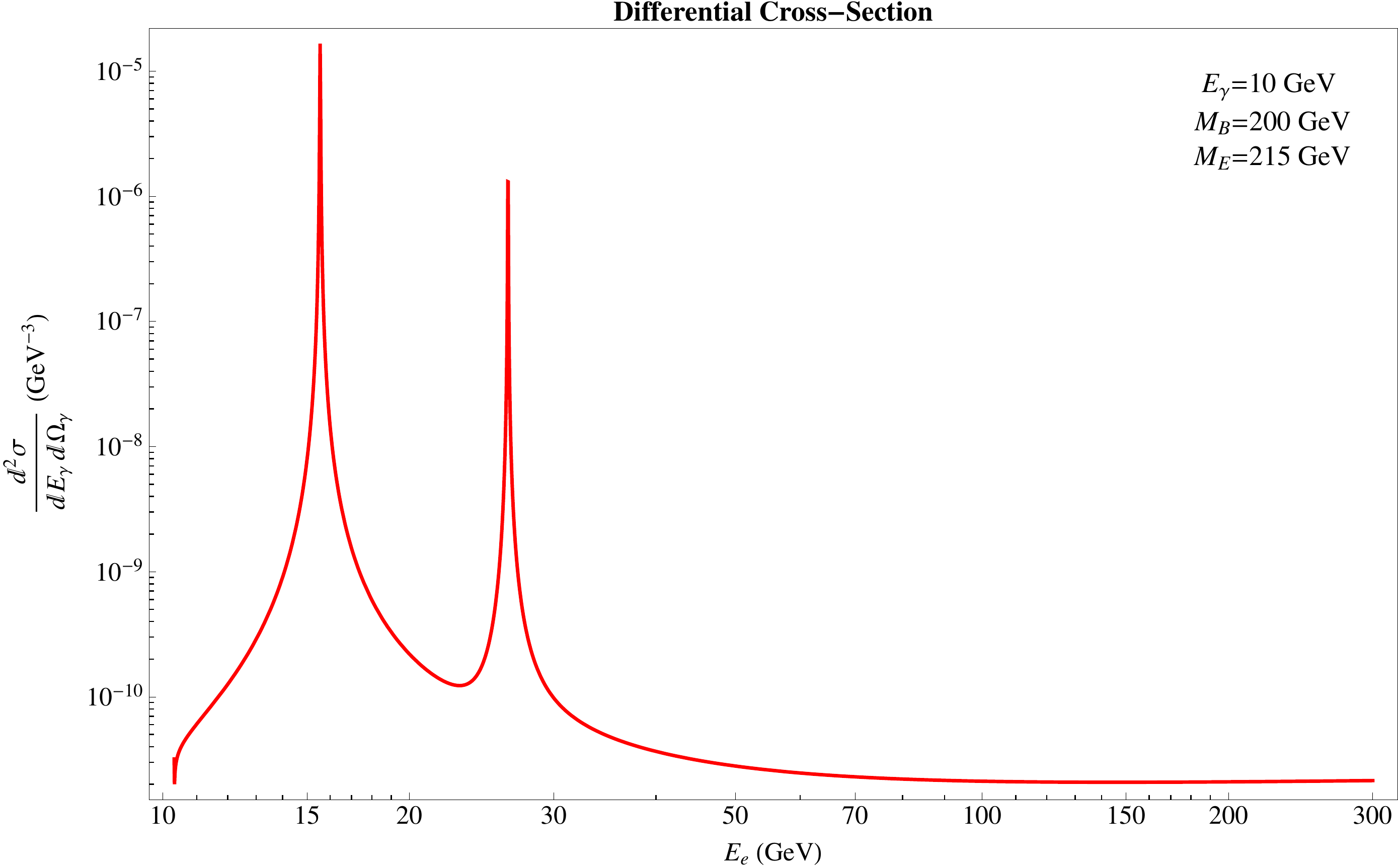}
\caption{The cross section for $B_H + e^- \to B_H + e^- + \gamma$ for photon energy $E_5 = 10$ GeV as a function of the incoming electron energy.  The resonance located near 15 GeV is due to pieces of the amplitude-squared which get contributions from diagrams $S_2$ and $S_3$ (which depend on the propagator term $\Sigma_{12}$), while the second resonance located near 27 GeV originates from diagrams $S_1$ and $S_3$ (which depend on the propagator term $\Sigma_{34}$).}
\label{fg:xn-vs-E2}
\end{figure}

The differential cross section as a function of the incoming electron's energy ($E_e = E_2$) is shown in Fig. \ref{fg:xn-vs-E2} for one particular photon energy ($E_\gamma = E_5 = 10$ GeV).  For this point in phase space, we obtain two resonances: the first (located near $M_E - M_B \simeq 15$ GeV) is due to pieces of the amplitude-squared which go like $\sim 1/| \Sigma_{12} |^2$, while the second resonance originates from pieces of the amplitude-squared which go like $\sim 1/| \Sigma_{34} |^2$.  We note that the first resonance always dominates over the second.

\begin{figure}[t]
\centering
\includegraphics[width=0.75\textwidth]{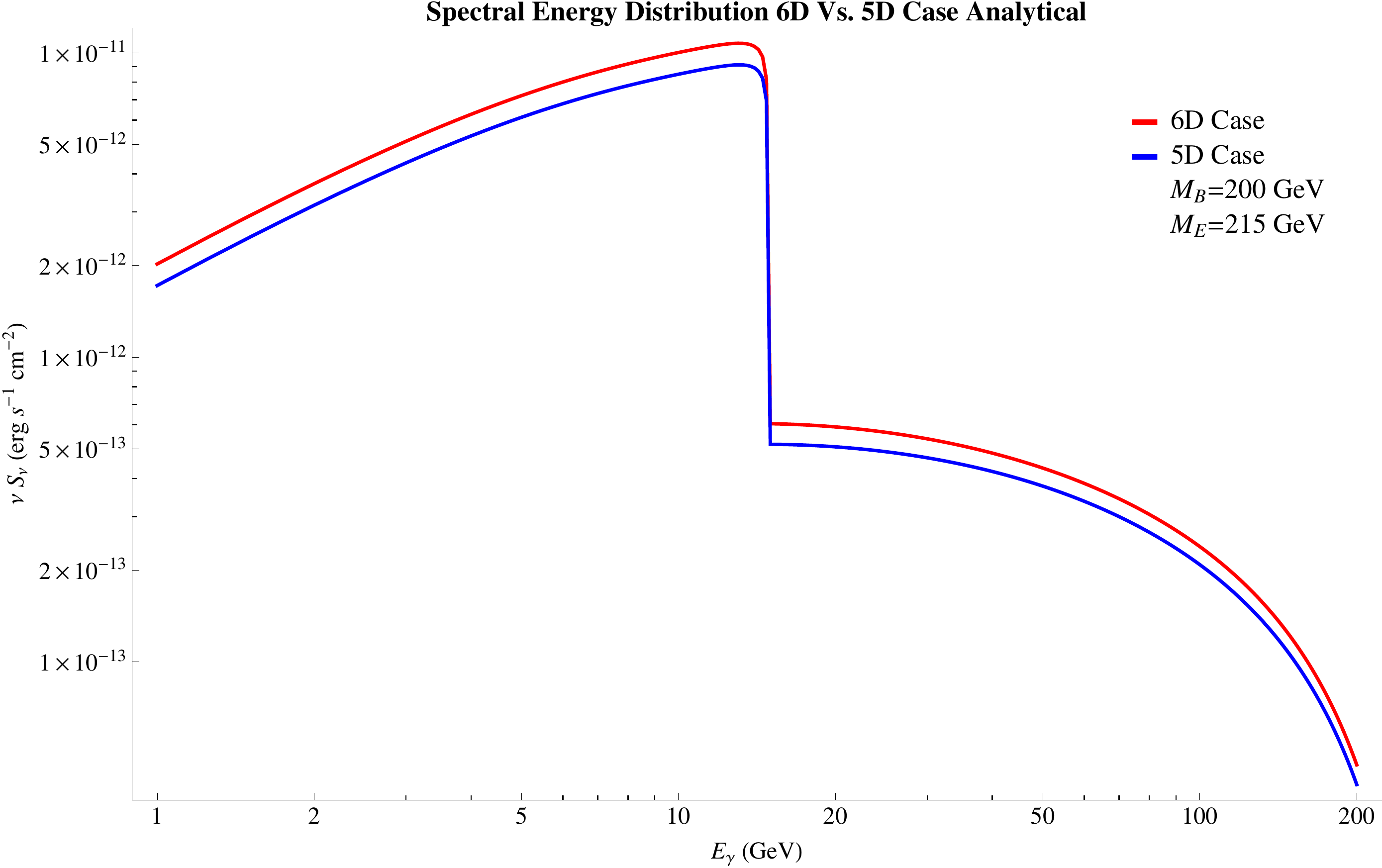}
\caption{The spectral energy distribution $\nu S_{\nu}$ ($= E_{\gamma}^{2} \times d\Phi_{\gamma}/dE_{\gamma}$) versus the photon energy from the interaction of WIMPs with the AGN jets of Centaurus A. Note that 5D case refers to a comparison to the model studied in \cite{Profumo:2010} and 6D case corresponds to the model studied in this work.}
\label{fg:flux-AGN-WIMP}
\end{figure}

We are now in a position to compute the gamma ray flux from interactions of AGN jet particles with halo WIMPs.  In the following, we will actually present our results in terms of the {\it spectral energy distribution}, $\nu S_\nu$, which is simply related to the flux as:
\begin{equation}
\nu S_\nu = E_\gamma^2 \times \frac{d\Phi_\gamma}{dE_\gamma} \,.
\end{equation}
Using Eqs. (\ref{eq:intlosdensitydm}), (\ref{eq:AGNjetfac}) and (\ref{eq:cross-sec}) in Eq. (\ref{eq:gammaraysflux}) and integrating over the initial electon's energy, we obtain spectra such as that shown in Fig. \ref{fg:flux-AGN-WIMP}.  For comparison, we separately calculated the results for the 5-d UED case which were previously done in Ref.~\cite{Profumo:2010}.  We note that, in both cases, the spectrum rises sharply as a function of the photon energy until a sharp cutoff at the mass difference between the WIMP and the up-scattered particle (i.e., the $E$ particle in the 6-d case).  The sharpness of this cutoff will be softened slightly once detector resolution effects are taken into account (see Section~\ref{sec:scan-results}).

Finally, we should note that we have performed this calculation in two ways.  In addition to calculating the photon flux using the {\it exact kinematics} as outlined above, we also computed the flux using the {\it collinear approximations} proposed in Ref. \cite{Profumo:2010}.  We found excellent agreement between the two calculations in the regions of the resonances and only small differences outside these regions.  Since the resonances give the largest contributions to the flux, we chose to use the more compact collinear-approximated calculation for our scans described in Section \ref{sec:scan-results} since these resulted in much faster computations.  The details of the cross section calculation in the collinear approximation are summarized in Appendix \ref{app:collinear}.

\section{Gamma Rays from Dark Matter Annihilation}
\label{sec:annihilations}

In this section, we compute the flux of gamma rays coming from WIMP annihilation.  In contrast to the flux from jet-halo WIMP interactions, the flux from WIMP annihilations depends quadratically on the density profile (because two WIMP particles are involved in the initial state) and is given by:
\begin{equation}
\left( \frac{d \Phi}{d E_\gamma} \right)_{ann.} = \frac{d N_\gamma}{d E_\gamma} \frac{\langle \sigma v \rangle_{tot}}{8 \pi M_B^2 d_{AGN}^2} \int_{r_{min}}^{r_0} dr 4 \pi r^2 \rho_{DM}^2 (r) \,,
\label{eq:ann-flux}
\end{equation}
where $\langle \sigma v \rangle_{tot}$ is the total annihilation cross section and $dN_\gamma/dE_\gamma$ is:
\begin{equation}
\frac{dN_\gamma}{dE_\gamma} = \frac{1}{\langle \sigma v \rangle_{tot}}\sum_f \langle \sigma v \rangle_f \frac{dN_\gamma^f}{dE_\gamma} \,.
\end{equation}
We use the index $f$ to denote the annihilation channels with one or more photons in the final state, $\langle \sigma v \rangle_f$ is the corresponding cross section and $dN_\gamma^f/dE_\gamma$ is the normalized photon spectrum per annihilation.

Below, we briefly outline the calculation of the two contributions to the WIMP annihilation gamma ray spectrum: ({\it i}) the continuum from annihilations into SM particles which then subsequently radiate or hadronize/decay into photons and ({\it ii}) line emissions from loop-induced decays directly into photons.  For more details, the interested reader should refer to Ref~\cite{Bertone:2009cb} where these calculations were first performed.

\subsection{Continuum}
\label{subsec:continuum}

As discussed earlier, pairs of LKPs annihilate predominantly into pairs of electroweak bosons $WW$ and $ZZ$, SM Higgs bosons $HH$ and (if heavy enough) top quark pairs $t\bar{t}$.  We compute the continuum gamma-ray spectrum from these annihilations using the micrOMEGAs code \cite{Belanger:2010gh}.  For $M_H = 125$ GeV and $M_B = 200$ GeV, the annihilation fractions are roughly 48\% $B_H B_H \to WW$, 28\% $B_H B_H \to HH$, 22\% $B_H B_H \to ZZ$ and  2\% $B_H B_H \to t\bar{t}$ and the total cross section is $\langle \sigma v \rangle = 2.40 \times 10^{-26}$ cm$^3$/s.

\begin{figure}[t]
\centering
\includegraphics[width=0.49\textwidth]{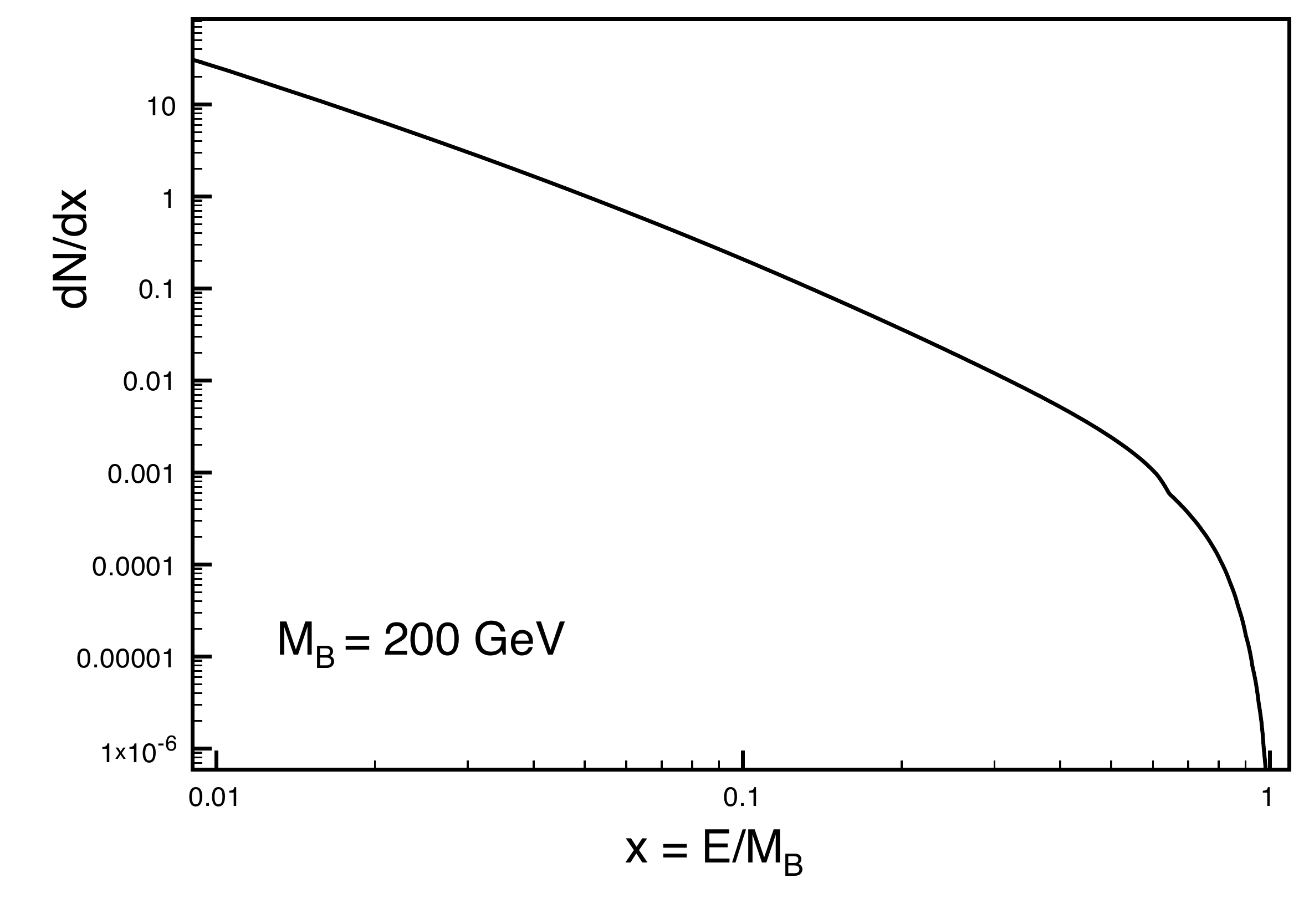}
\includegraphics[width=0.49\textwidth]{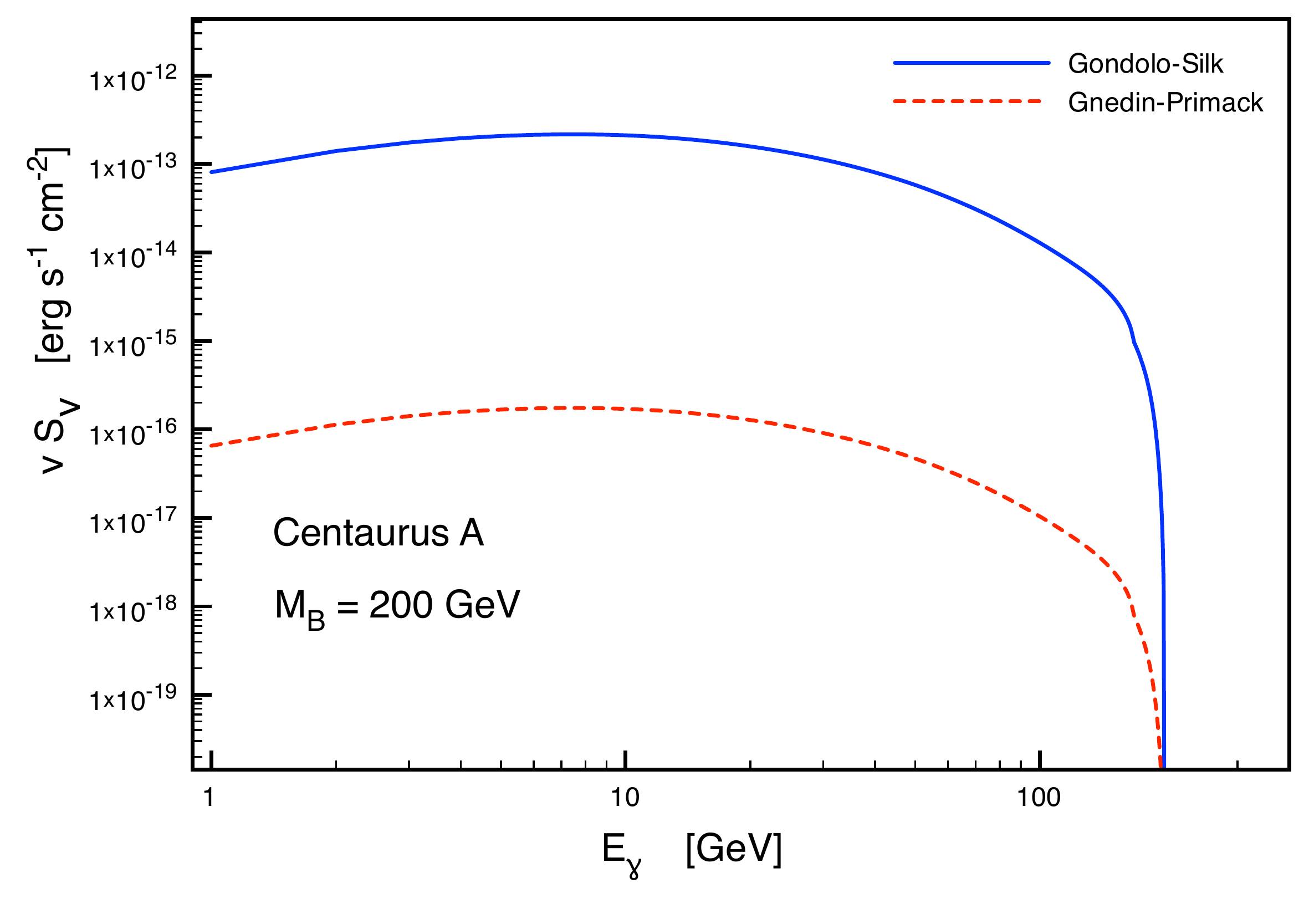}
\caption{Gamma ray spectra from continuum WIMP annihilations.  Left: The differential photon spectrum per annihilation event as a function of the fractional photon energy $x = E_\gamma/M_B$ for $M_B = 200$ GeV. Right: The spectral energy distribution $\nu S_\nu$ ($= E_\gamma^2 \times d\Phi_\gamma/dE_\gamma$) versus the photon energy for Centaurus A for two choices of the dark matter density profile.}
\label{fg:dNdx}
\end{figure}

In Fig.~\ref{fg:dNdx} (left), we plot the differential photon spectrum $dN/dx$ where $x = E_\gamma/M_B$.  In Ref. \cite{Bertone:2009cb}, it was pointed out that, in contrast to other models, the continuum spectrum from the chiral square model sharply decreases well before the value of the WIMP mass $M_B$.  The reason is that the main annihilation modes are dominated by ``photon unfriendly'' modes, consisting of massive (and often neutral) particles which are unlikely to radiate high-energy photons.  The result is that the bulk of photons come from radiation (or after hadronization, decays of $\pi^0$'s) from the even softer decay products of the particles produced in the primary annihilation.

\begin{figure}[t]
\centering
\includegraphics[scale=0.4]{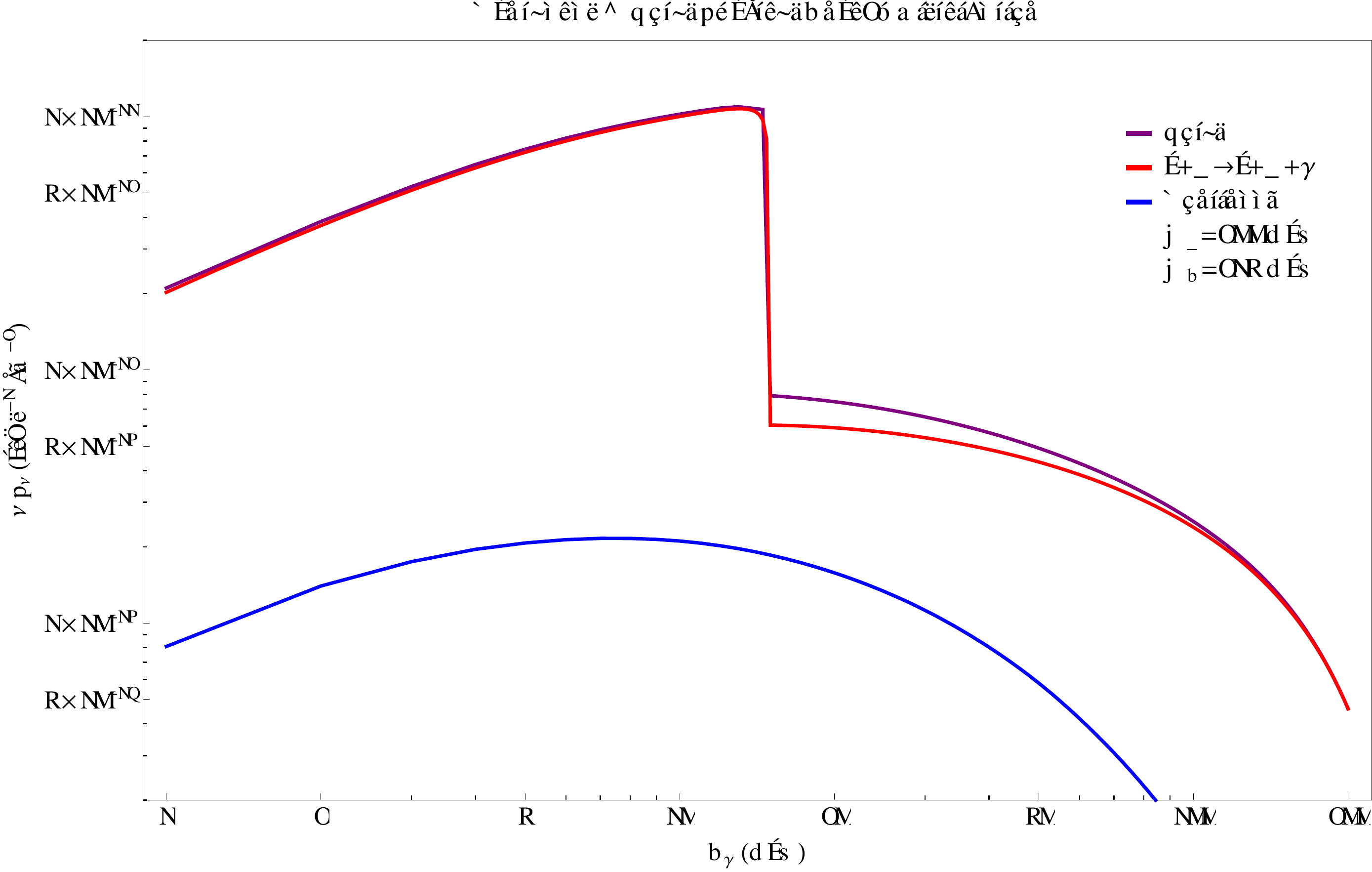}
\caption{Total spectral energy distribution for the process $e+B$ with one or two final photons (leading order and one loop) including continuum emission from dark matter annihilation.}
\label{fg:flux-AGN-WIMP-total}
\end{figure}

Fig.~\ref{fg:dNdx} (right) depicts the spectral energy distribution for continuum gamma rays from WIMP annihilations for Centaurus A using two different dark matter density profiles (the Gondolo-Silk profile discussed above and the Gnedin-Primack profile of Ref. \cite{Gnedin:2003rj}).  Comparing to the spectrum from jet-halo interactions, we see that below $\sim$ 15 GeV the jet-halo gamma rays dominate by an order of magnitude, but above this value of $E_\gamma$ the continuum from WIMP annihilations is of the same order as the jet-halo contribution.  Fig.~\ref{fg:flux-AGN-WIMP-total} shows the sum of the jet-halo and continuum annihilation gamma ray spectra for a particular choice of $M_B$ and the mass splitting ($M_E - M_B$).

\subsection{Line Emission}
\label{subsec:line-emission}

Let us now consider the direct (loop-level) annihilations into gamma rays, $\gamma + X$.  In the chiral square model, the WIMP is a scalar particle such that the only possibility is for $X$ to be a vector particle.  This allows for the possibility of three lines originating from the $\gamma \gamma$, $\gamma Z$ and $\gamma B^{(1,1)}$ final states (where $B^{(1,1)}$ is the (1,1) KK excitation of the hypercharge gauge boson with mass roughly $\sim \sqrt{2} M_B$).  In fact, the chiral square model has been shown to produce a gamma ray spectrum from our galactic center that possesses the rare characteristic of multiple {\it and} well-separated line features (even after the inclusion of detector resolution effects) \cite{Bertone:2009cb}.

\begin{figure}[t]
\centering
\includegraphics[scale=0.3]{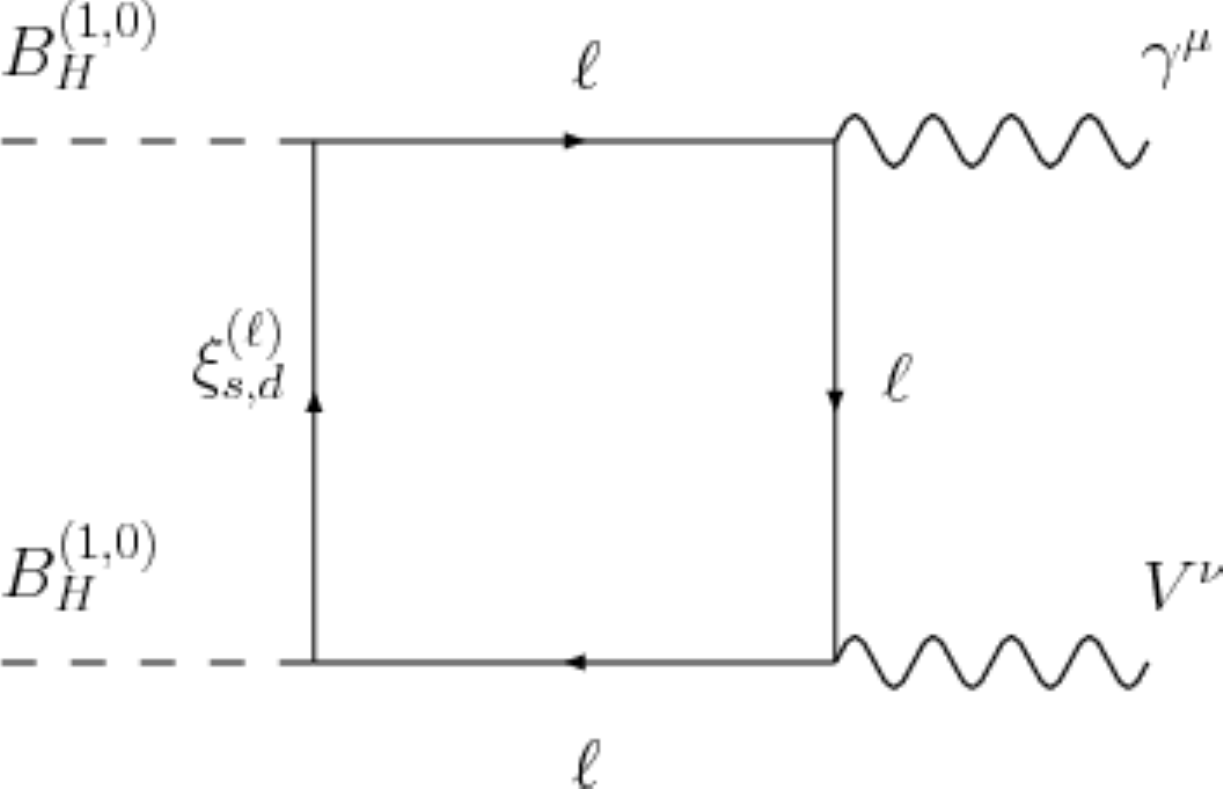} \hspace{0.25 cm}
\includegraphics[scale=0.3]{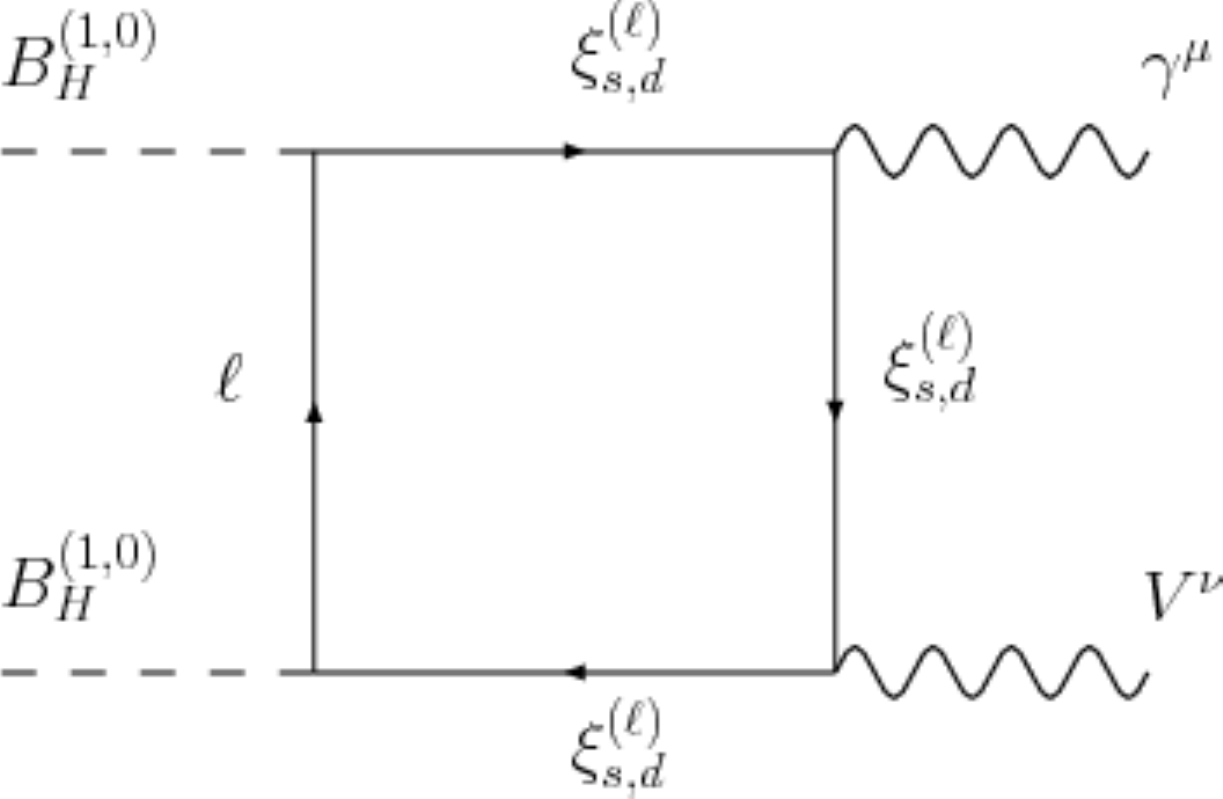} \hspace{0.25 cm}
\includegraphics[scale=0.3]{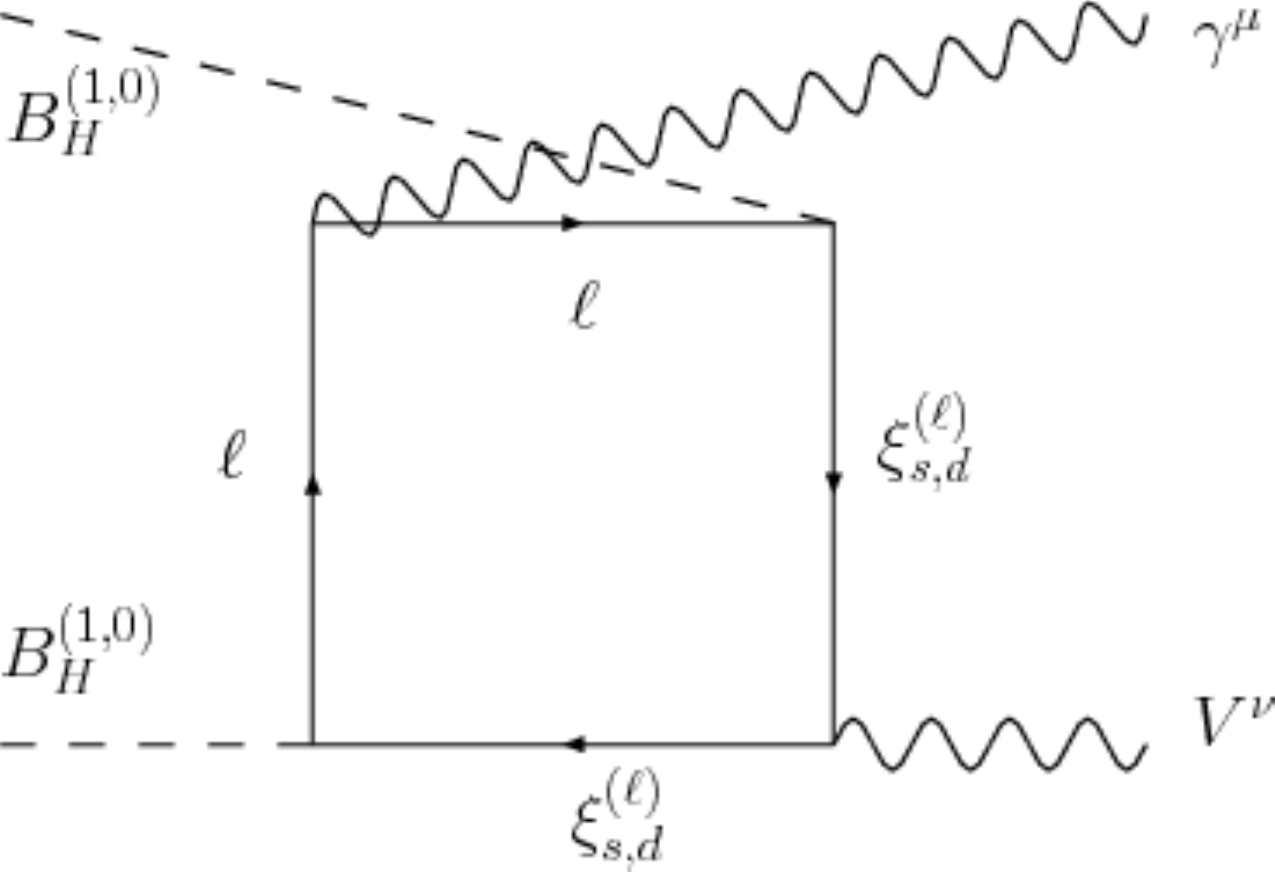}
\caption{Example Feynman diagrams for the process $B_H + B_H \to \gamma + V$.  $\xi^{(\ell)}_{s,d}$ are the gauge eigenstates for the (1,0) excitations of the SM leptons ($\ell$).}
\label{fg:loop-diagrams}
\end{figure} 

\begin{figure}[t]
\centering
\includegraphics[scale=0.45]{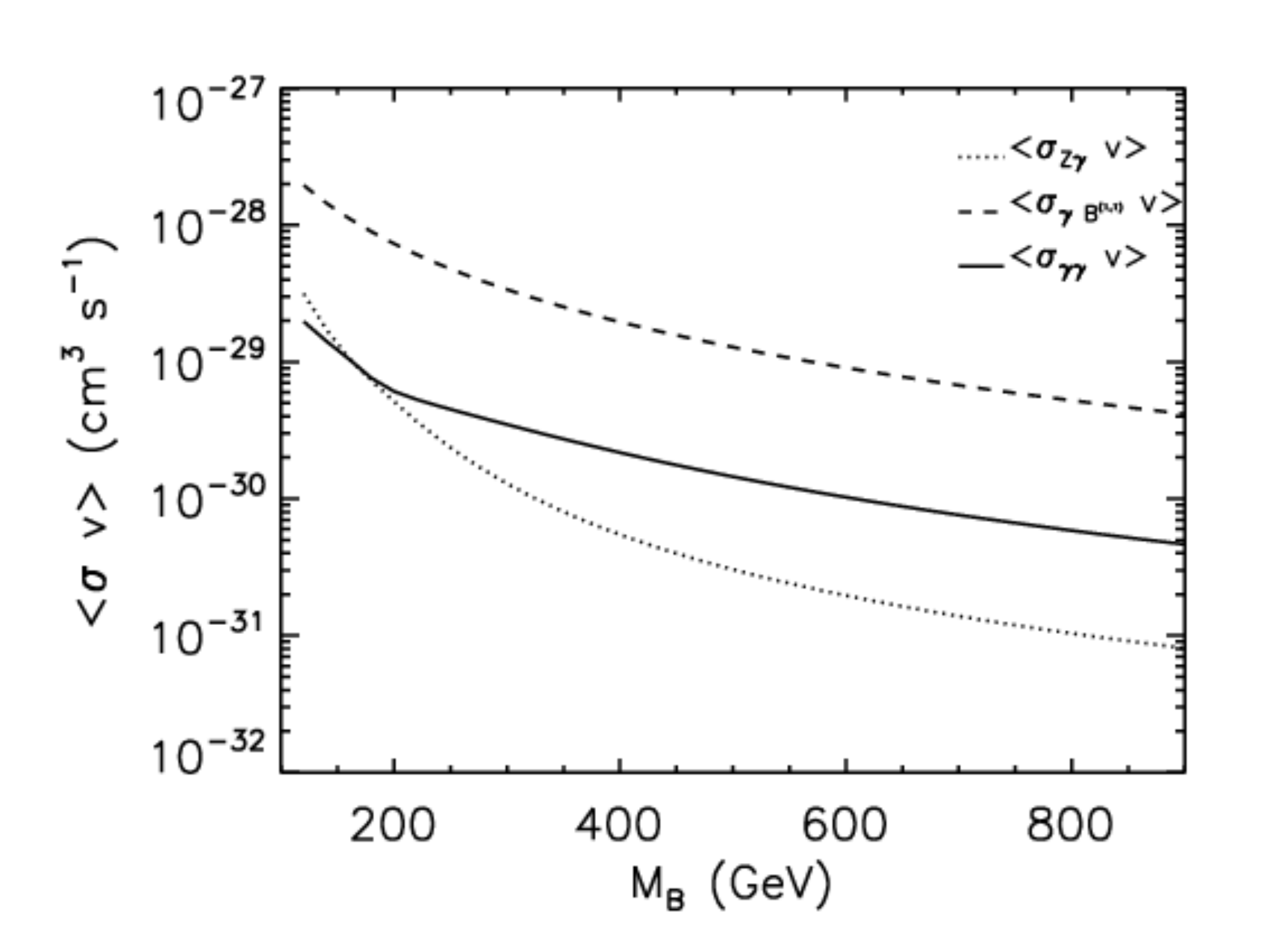}
\caption{The cross sections for $B_H B_H \to \gamma \gamma, \gamma Z$ and $\gamma B^{(1,1)}$ as a function of the WIMP mass. This plot was taken from Ref. \cite{Bertone:2009cb}.}
\label{fg:Sigma-loop}
\end{figure}

Sample Feynman diagrams which contribute to $B_H(p_1) B_H(p_2) \to \gamma(p_A) V(p_V)$ are depicted in Fig. \ref{fg:loop-diagrams}.  The amplitude for this process can be written as:
\begin{equation}
{\cal M} = \epsilon^{\mu *}_A(p_A) \epsilon^{\nu *}_B(p_B) {\cal M}^{\mu\nu}(p_1,p_2,p_A,p_B) \,,
\label{eq:loop-amplitude}
\end{equation}
where $\epsilon_A^\mu$ and $\epsilon_B^\nu$ are the polarization tensors of the photon and $V$ gauge boson, respectively. In general, the tensor ${\cal M}^{\mu\nu}$ can be expanded in terms of the metric tensor and external momenta as:
\begin{eqnarray}
{\cal M}^{\mu\nu} &=& A g^{\mu\nu} + B_1 p_1^\mu p_1^\nu + B_2 p_2^\mu p_2^\nu + B_3 p_1^\mu p_2^\nu + B_4 p_2^\mu p_1^\nu + B_5 p_B^\mu p_A^\nu \nonumber\\ 
&&+ B_6 p_1^\mu p_A^\nu + B_7 p_B^\mu p_1^\nu + B_8 p_2^\mu p_A^\nu + B_9 p_B^\mu p_2^\nu \,.
\end{eqnarray}
It was shown in Ref.~\cite{Bertone:2009cb}, that using the non-relativistic nature of the WIMPs (i.e., $p_1 \simeq p_2 \equiv (M_B, {\bf 0})$) along with conservation of momentum, the only surviving term is $g^{\mu\nu}$ and, thus, the computation of the cross section is very much simplified.  Fig. \ref{fg:Sigma-loop} shows the cross sections for $\gamma\gamma$, $\gamma Z$ and $\gamma B^{(1,1)}$ as a function of $M_B$ for the KK masses $M_E = 1.2 M_B$ and $M_{B^{(1,1)}} = 1.6 M_B$.

The enhancement of the $\gamma B^{(1,1)}$ cross section with respect to the other two may seem surprising due to the fact that the $B^{(1,1)}$ is so much more massive than the photon or $Z$.  However, the effect is understood as follows.  In the case where $V$ is either a photon or a $Z$ boson, the couplings between $V\ell \bar{\ell}$ and $V\xi \bar{\xi}$ (where $\xi$ are the gauge eigenstates of the (1,0) excitations of the SM leptons) are nearly the same strength and the result is a significant cancellation between the various diagrams.  However, in the case where $V$ is identified with the $B^{(1,1)}$ KK gauge boson, the $V\ell \bar{\ell}$ couplings are loop-suppressed, while the $V\xi \bar{\xi}$ couplings are relatively large.  This results in less cancellation between the various diagrams and an enhanced cross section compared to the other two processes.

The spectra for the lines will depend on the intrinsic width of the particle produced in association with the photon.  In other words, while the $\gamma\gamma$ line will simply be a delta function at the WIMP mass $M_B$, lines arising from the process $B_H B_H \to \gamma V$ will possess an intrinsic width which will depend on the mass of the boson in the final state $M_V$:
\begin{equation}
\frac{dN^V_\gamma}{dE} = \frac{ 4 M_B M_V \Gamma_V}{f_1 f_2} \,,
\end{equation}
where $\Gamma_V$ is the width of $V$ and:
\begin{eqnarray}
f_1 &\equiv& \left[ \tan^{-1} \left( \frac{M_V}{M_B} \right) + \tan^{-1} \left( \frac{4 M_B^2 - M_V^2}{M_V \Gamma_V} \right) \right] \, \\
f_2 &\equiv& \left[ \left( 4 M_B^2 - 4 M_B E_\gamma - M_V^2 \right)^2 + \Gamma_V^2 M_V^2 \right] \,.
\end{eqnarray}

In practice, the detector resolution tends to dominate the line shape for the cases we consider.

\section{Results}
\label{sec:scan-results}

In this section, we combine the fluxes from jet-halo WIMP interactions with those of WIMP annihilations (both continuum and line emissions) and compare with the data collected by the Fermi gamma ray telescope \cite{Falcone:2010fk}.  In addition to these fluxes, we add a background contribution assumed to originate from astrophysical processes that behaves as a power law:
\be
{d\Phi_{\rm bkg}\over dE_\gamma} = A_{\rm b} \left({E_\gamma\over{\rm GeV}}\right)^{\delta_{\rm b}} \,.
\ee
We assume the $\gamma$-ray spectra is unaltered in route to measurement and we smear the energy according to a Gaussian kernel:
\be
{d\Phi_{\rm tot}\over dE_\gamma} =G_0(E^\prime_\gamma,E_\gamma)  \left({d\Phi_{\rm bkg}\over dE^\prime_\gamma}+{d\Phi_{\rm cont.}\over dE^\prime_\gamma}+{d\Phi_{\rm line}\over dE^\prime_\gamma}+\lambda_{\rm AGN}{d\Phi_{\rm AGN}\over dE^\prime_\gamma}\right) dE^\prime_\gamma \,,
\ee
where the kernel is:
\be
G_0(E^\prime_\gamma,E_\gamma) = {1\over \sqrt{2\pi} \sigma_{exp}} e^{-{(E^\prime_\gamma-E_\gamma)^2\over 2 \sigma_{exp}^2}} \,.
\ee
$\sigma_{exp}$ is the experimental resolution, which we adopt as $\sigma_{exp} = 0.1 E_\gamma^\prime$ for Fermi LAT.  The scale factor $\lambda_{\rm AGN}$ parameterizes the uncertainty in the AGN contribution, which we take to be $0.5-1$.  To arrive at the $\chi^2$ contribution from the AGN data, we average the event rate over the energy bins.  

To fit the model, we perform a Bayesian fit to the data, the details of which are provided in Appendix~\ref{apx:mcmc}.  The fit includes the Centaurus-A AGN data~\cite{Falcone:2010fk}.  We fit the dark matter relic abundance in one of two ways:
\be
\Omega_{B_H} h^2 = 0.11 \pm 0.01,
\ee
which we denote as the ``saturated'' case, where the relic abundance is explained entirely by the dark matter candidate.  Here, we take a 10\% uncertainty that is prescribed to be the theory uncertainty on top of the excellent experimental measurements of Planck~\cite{Ade:2013zuv}.  In the other case, we consider only the measurement as an upper bound of the dark matter relic abundance which we refer to as the ``unsaturated'' case.  We will see that these two circumstances lead to different characteristic spectra, and therefore a different overall fit to the AGN data.  

To further constrain the model, we will also include in our fit the most current limits on WIMP parameter space from direct detection experiments (in particular, those from Xenon100~\cite{Aprile:2012nq}).  Since we allow for a reduced relic abundance, we scale the scattering rate exclusion and $B_H$ annihilation observables with the relic abundance:
\be
\sigma^{\rm meas.}_{\rm SI} =\sigma_{\rm SI} {\Omega_{B_H} h^2 \over \Omega_{\rm meas.} h^2},\quad\quad \sigma v^{\rm meas.}_{\gamma X} = \sigma v_{\gamma X} \left({\Omega_{B_H} h^2 \over \Omega_{\rm meas.} h^2}\right)^2,
\ee
where the measured rate must be compatible with present experimental bounds from Fermi LAT~\cite{fermi_line2012} and Xenon100~\cite{Aprile:2012nq}.

\begin{figure}[h]
\begin{center}
      \includegraphics[angle=0,width=0.49\textwidth]{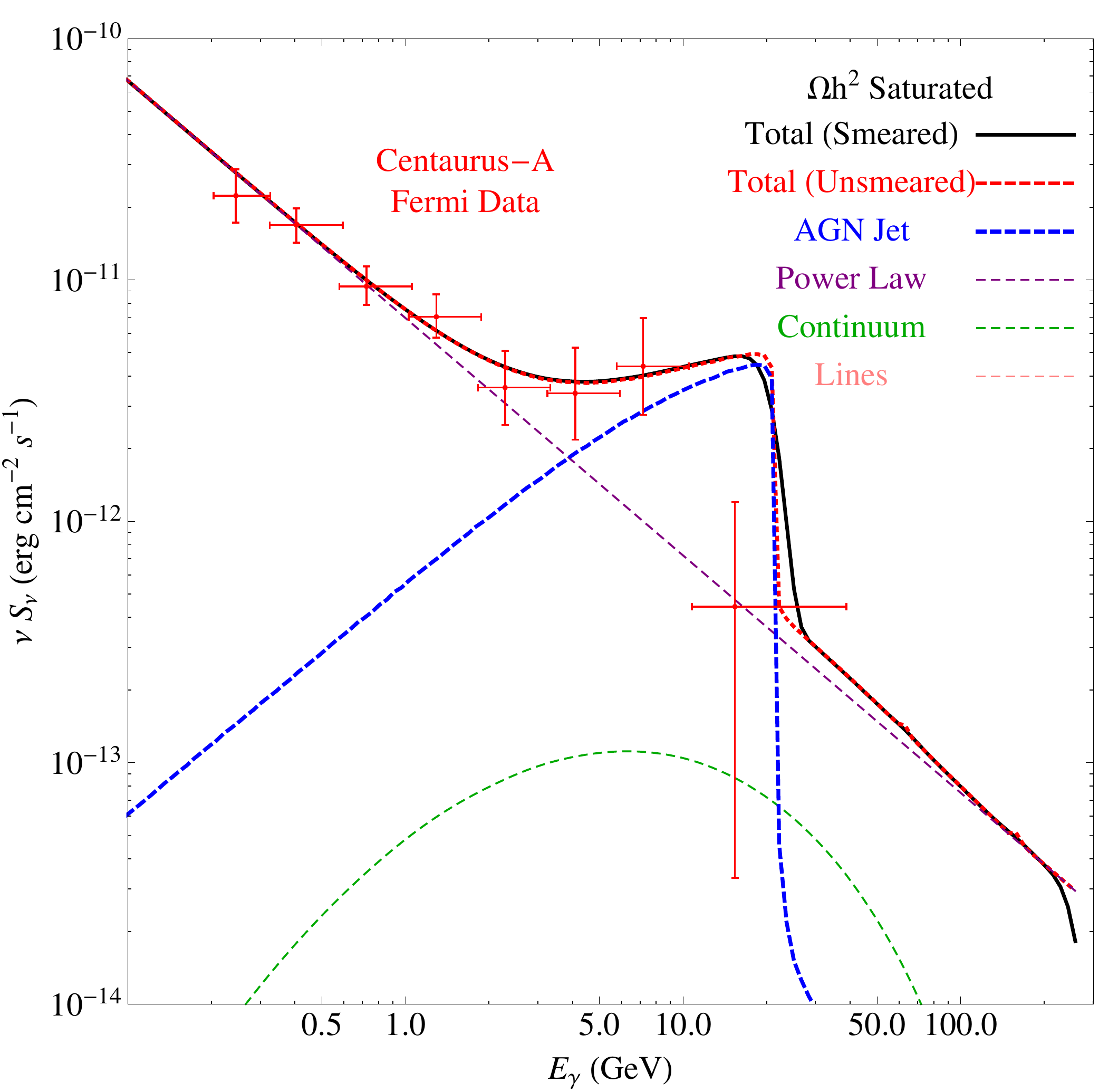}
      \includegraphics[angle=0,width=0.49\textwidth]{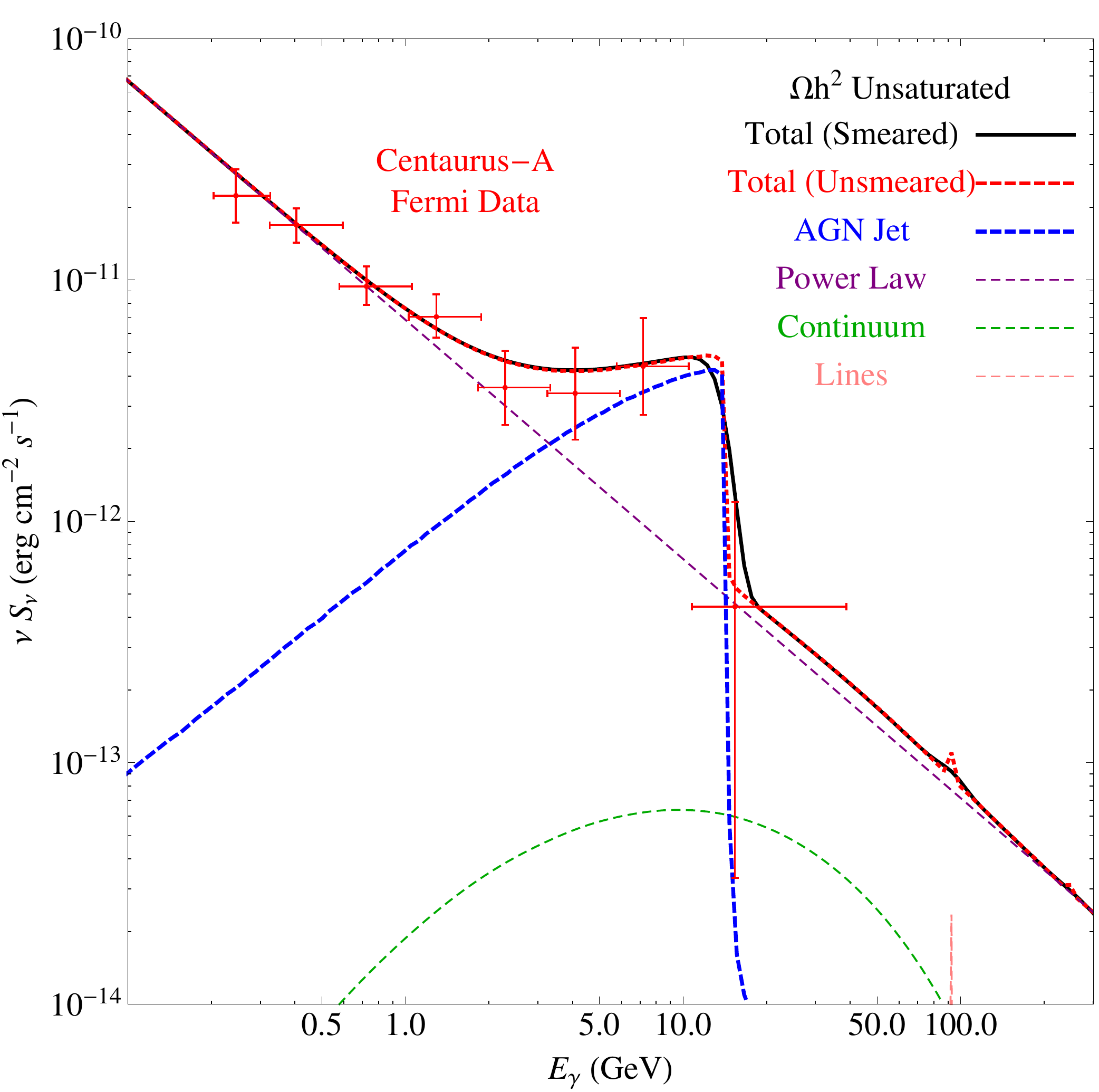}
\caption{Photon spectra from the best fit point for the saturated (left) and unsaturated (right) relic abundance scenarios. The unsaturated relic abundance case yields an energy cutoff that fits the Centaurus-A data slightly better than the saturated case. While the continuum and lines are generally suppressed below the astrophysical background, the $B^{(1,1)}\gamma$ line can appear well above the continuum, offering possible detection in our own galaxy.}
\label{fig:AGN}
\end{center}
\end{figure}
In Fig.~\ref{fig:AGN}, we show the $\gamma$-ray spectra from the best-fit of the Centaurus-A data for the two relic abundance scenarios we consider.  For the astrophysical background, we find the power law parameters for the saturated case to be $A_{\rm b} = (6.1^{+1.0}_{-0.8})\times10^{-12}$ and $\delta_{\rm b} = -3.06 \pm 0.14$, with the unsaturated scenario having a very similar fit.
We find that the saturated relic abundance case yields a cutoff that is of higher $E_\gamma$ than in the unsaturated case.  We observe that since this cutoff is related to the mass difference
\be
E_\gamma^{\rm cut} = M_E - M_{B},
\ee
the model is forced into a particular and well defined relic density scenario which we will discuss below.  Generally, the continuum from dark matter annihilation and the $\gamma$-ray lines are suppressed well below the astrophysical background.  However, the $B^{(1,1)}\gamma$ line can appear well above the continuum, allowing detection within our own galaxy.

\begin{figure}[!htpb]
\begin{center}
      \includegraphics[angle=0,width=0.47\textwidth]{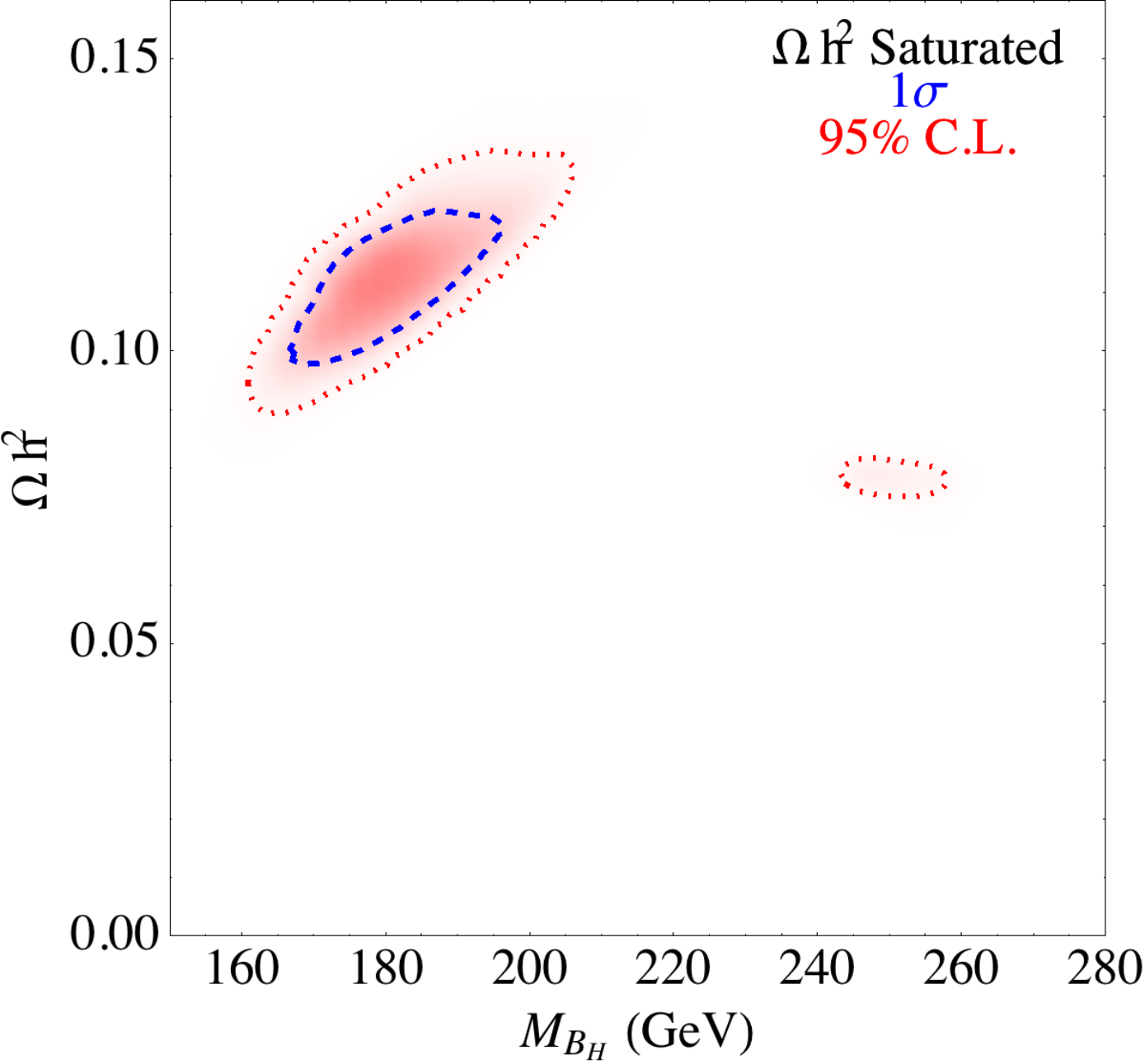}
      \includegraphics[angle=0,width=0.47\textwidth]{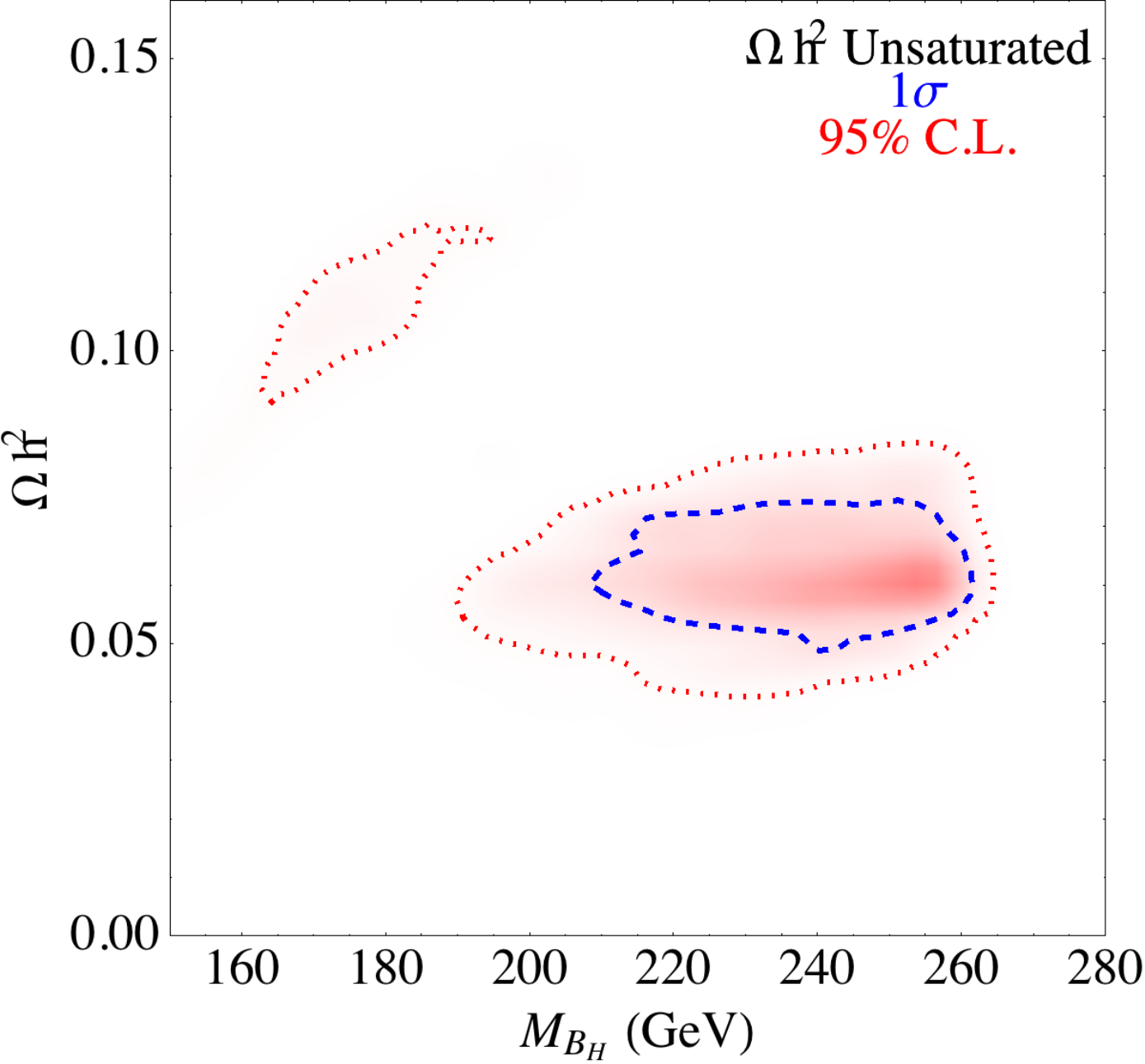}\\
      \includegraphics[angle=0,width=0.47\textwidth]{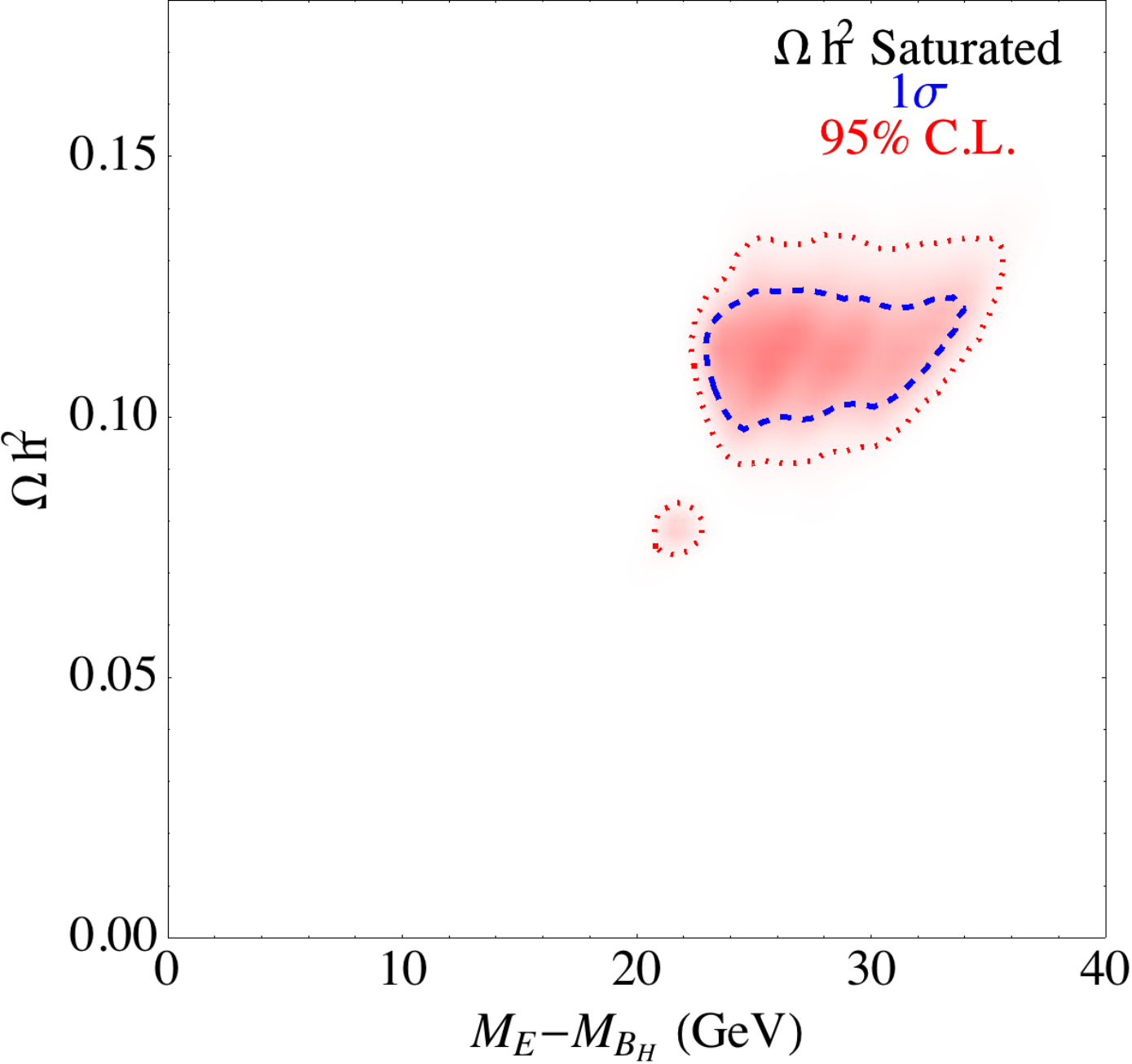}
      \includegraphics[angle=0,width=0.47\textwidth]{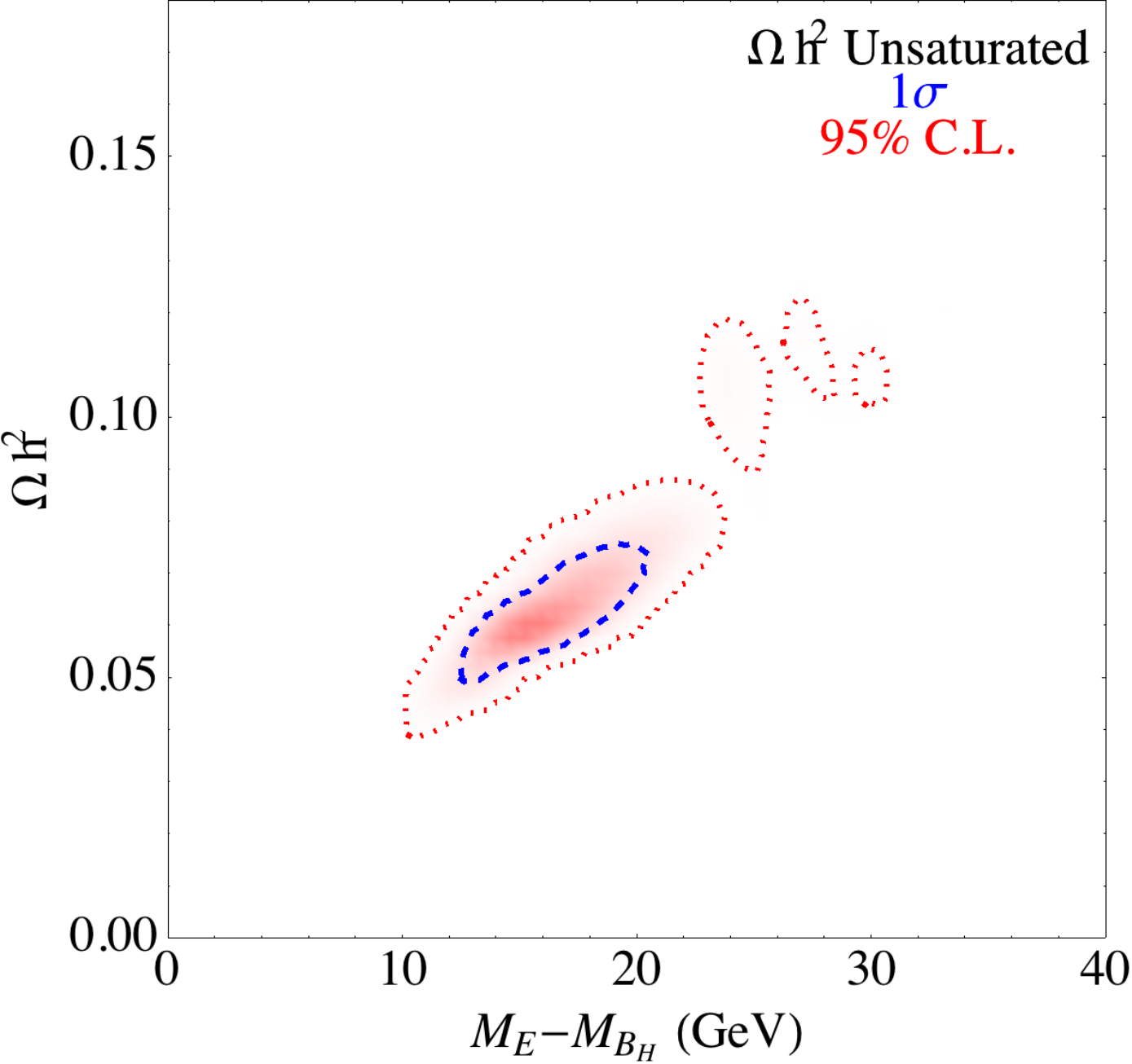}
\caption{Dark matter relic abundance and mass posterior distributions while requiring the DM candidate is not above the measured abundance (top left panel), and fully saturates the measured abundance (top right panel).  The bottom panels show the relic abundance and mass splitting illustrating the role of $B_H-E$ co-annihilation. The dashed (dotted) contour contains the $1\sigma$ (95\%) C.L. regions, respectively.}
\label{fig:rd}
\end{center}
\end{figure}
In the top panels of Fig.~\ref{fig:rd}, we show 2-d posterior distribution of the relic density and dark matter mass under the two above scenarios.  In the unsaturated case, there is a preferred region at nearly half that of the measured relic abundance with mass in the $180-260$ GeV range.  However, in the other case, the relic density is indeed saturated, but with a lower dark matter mass of $160-200$ GeV.  The source of the lower relic abundance in the first case is due to coannihilations of $B_H$ with the compressed spectrum of the first excitations.  This is highlighted in bottom panels of  Fig.~\ref{fig:rd}, which shows the relic abundance and mass splitting of the excited electron.  As previously seen, this is important as the ``edge'' in the Fermi data point to a mass splitting on the order of $15$ GeV.  A spectrum with a mass splitting this small has a significant coannihilation rate, thereby lowering the relic abundance.  Alternatively, a non-standard cosmological scenario such as non-thermal mechanisms that enhance the relic abundance must be realized.

\begin{figure}[!htpb]
\begin{center}
      \includegraphics[angle=0,width=0.49\textwidth]{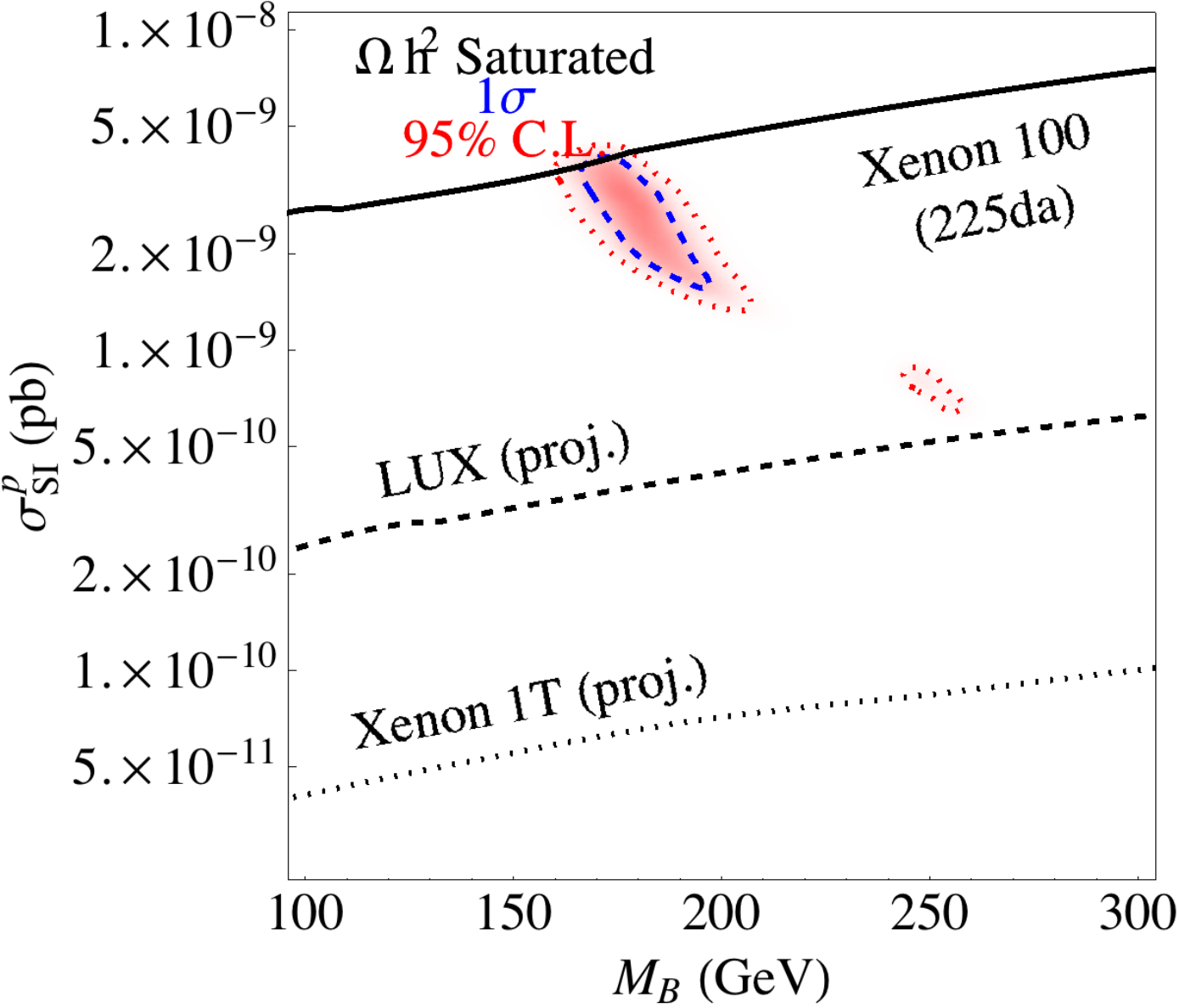}
      \includegraphics[angle=0,width=0.49\textwidth]{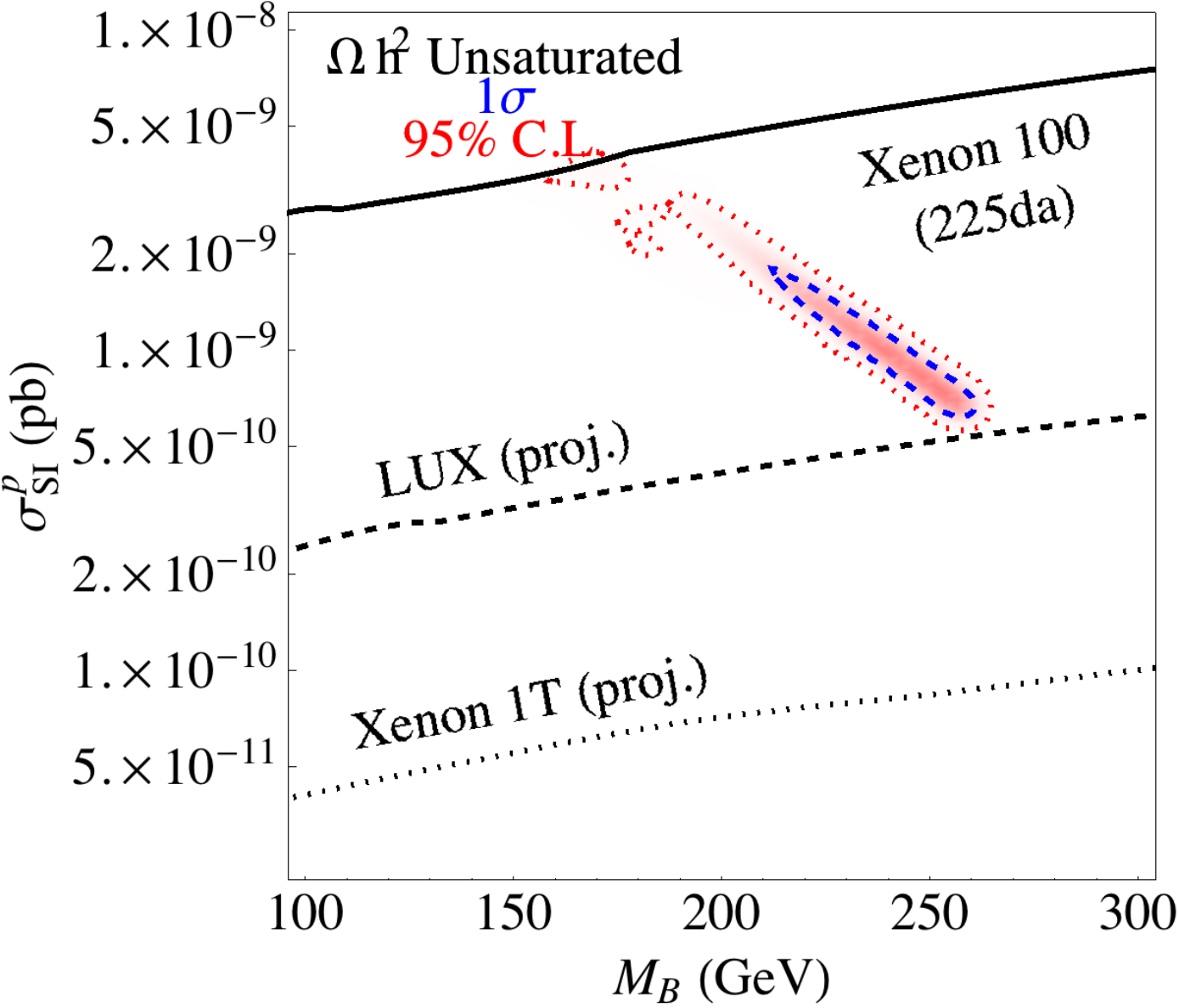}
\caption{Scattering cross section for dark matter mass in the unsaturated (left) and saturated (right) relic abundance cases. The dashed (dotted) contour contains the $1\sigma$ (95\%) C.L. regions, respectively.  Either scenario is easily probed by the LUX and Xenon 1T experiments.}
\label{fig:sip}
\end{center}
\end{figure}
\begin{figure}[~htpb]
\begin{center}
      \includegraphics[angle=0,width=0.49\textwidth]{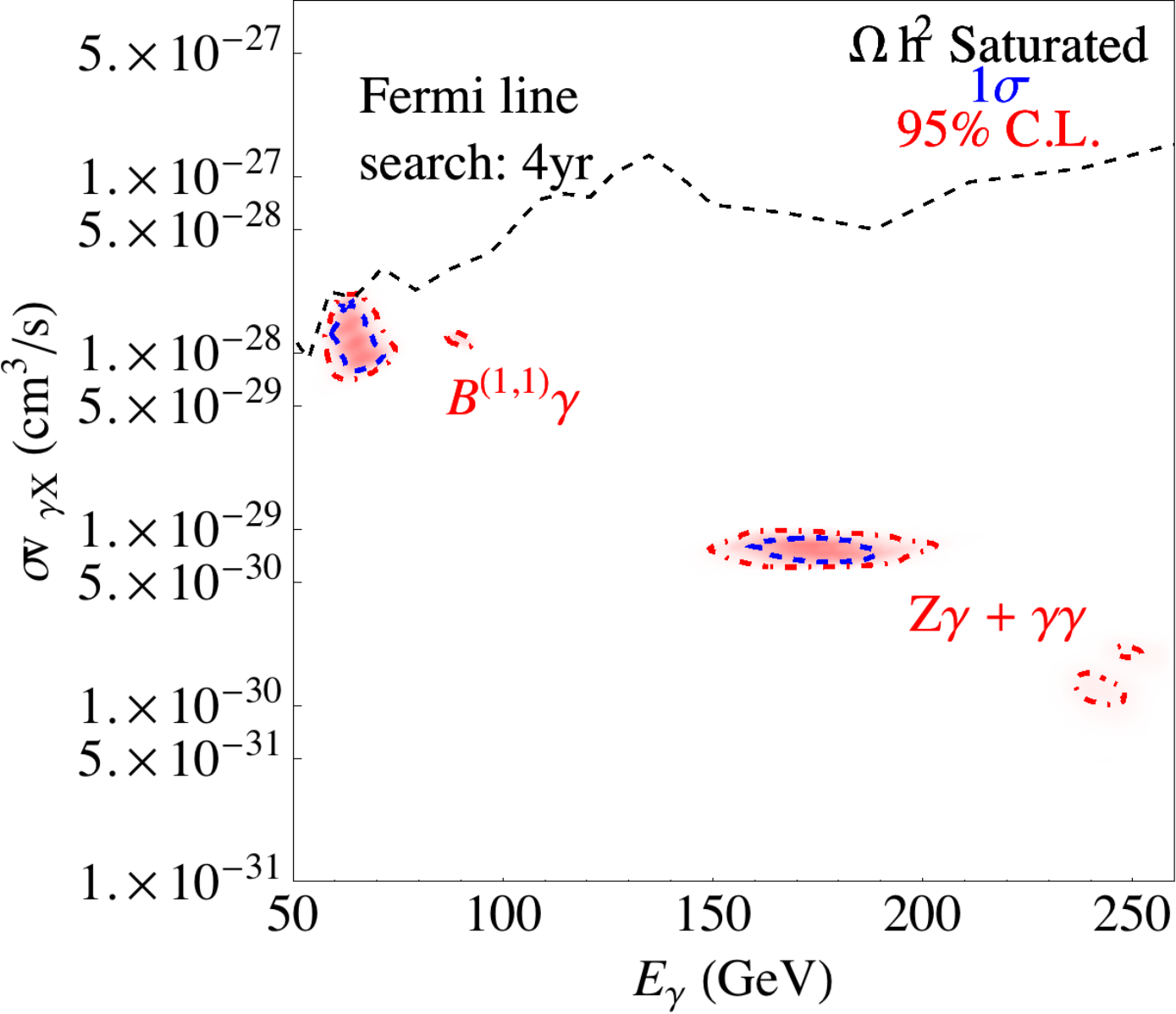}
      \includegraphics[angle=0,width=0.49\textwidth]{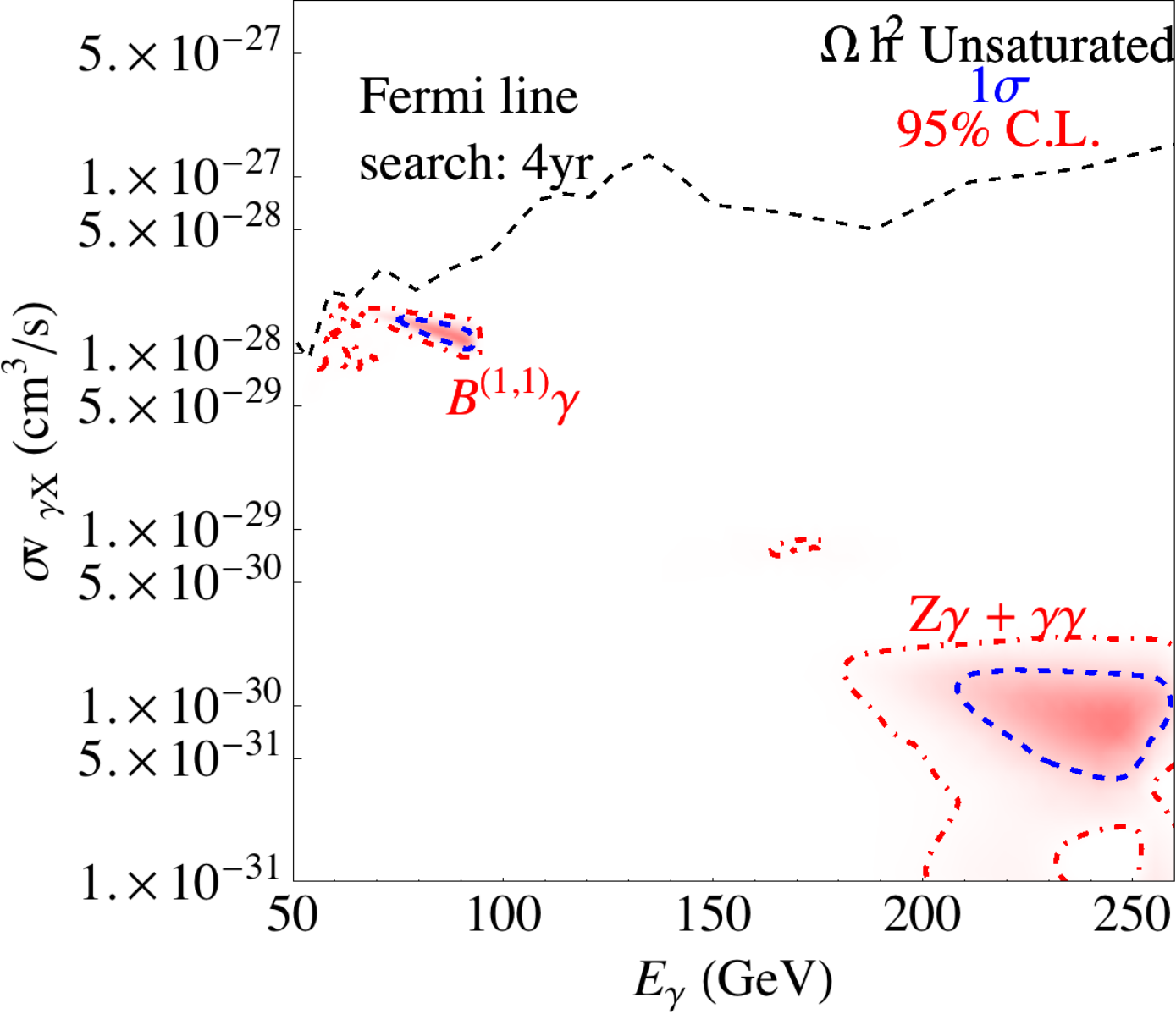}
\caption{Cross section distribution of $\gamma$-line from $B_H B_H \to X+\gamma$ with and without the relic density saturated. The dashed (dotted) contour contains the $1\sigma$ (95\%) C.L. regions, respectively.  The line cross section limit from Fermi after 4 years is given by the dashed line above the contours. }
\label{fig:lines}
\end{center}
\end{figure}

Finally, in Figs.~\ref{fig:sip} and \ref{fig:lines}, we present associated dark matter observables.  In Fig.~\ref{fig:sip}, we see the spin-indpendent scattering cross section for the two relic abundance cases.  In the case of a saturated relic abundance, the expected cross sections are quite close to the Xenon100 limit~~\cite{Aprile:2012nq}.  This indicates some amount of tension within the model.  However, the unsaturated case is naturally within a factor of 3-4 of the current bound.  Either case is within reach of future direct detection experiments such as LUX~\cite{Akerib:2012ak} and Xenon-1Ton~\cite{xenon1t} with a cross section no lower than ${\cal O}(5\times 10^{-10}~{\rm pb})$.  

In Fig.~\ref{fig:lines}, we show the location and predicted cross section into the $\gamma$-ray lines $\gamma\gamma$, $Z\gamma$ and $B^{(1,1)}\gamma$ and the present 4 year Fermi line search limit~\cite{fermi_line2012}.  The $B^{(1,1)}\gamma$ is sizable and is well positioned for Fermi line searches within our own galaxy in the near future.  We find the $B^{(1,1)}\gamma$ should have a line location of $50-100$ GeV and cross section at the ${\cal O}(10^{-28}~{\rm cm}^2/{\rm s})$ level.  The $\gamma\gamma$ and $Z\gamma$  lines are of similar strength.  Furthermore, given the energy resolution of $\gamma$-ray experiments, they will likely be smeared together as one line at this energy.  Therefore, they are combined in the contour plot as one line.

\section{Discussion}
\label{sec:discussion}

Given the results of the fit to the Fermi data for Centaurus-A, we now discuss further implications consistent with these data.  Since the best fit points in the model correspond to a mass splitting of ${\cal O}(15~{\rm GeV})$, we anticipate that the relic density of $B_H$ should not necessarily be the value measured by WMAP and Planck due to coannihilation with the up-scattered state, $E$.  Therefore, a second dark matter candidate must be needed to explain the remaining dark matter of the universe; a simple explanation is the axion arising from the Peccei-Quinn solution of the strong CP problem.  

With a small mass splitting between the dark matter candidate and the compressed spectrum associated with the first level excitation, we expect searches at the LHC to have difficulty discovering this scenario.  This is due to the soft decay products that fall below the cuts required in such analyses.  Indeed, the null searches for new heavy states in models such as supersymmetry have not precluded this scenario.  However, a mass scale of ${\cal O}(200~{\rm GeV})$ is within reach of a future ILC at 250 or 500 GeV.  There, the cleaner background and more precise control over initial state energy may allow a discovery.

It is worth noting that in the presence of brane kinetic terms, the masses in extra dimension models are allowed to vary more freely.  In this case, while the strength of coannihilation between $B_H$ and the compressed spectrum will be weakened, we can say that coannihilation between $B_H$ and $E$ will take place to some level due to the AGN jet data requiring a $10-20$ GeV mass splitting.  However, the searches at the LHC may uncover the decay of, say, $Q^{(1,0)}\to B_H q$ with a sufficient mass splitting.

Future data from many experiments should further probe this scenario.  The Fermi LAT should be able to shortly discover the $B^{(1,1)}\gamma$ line. Moreover, future direct detection experiments such as LUX~\cite{Akerib:2012ak} and Xenon-1Ton~\cite{xenon1t} are well positioned to probe the entire scattering cross section, assuming the $B_H$ relic abundance is not suppressed too much.  The Fermi LAT and HESS measurements on Centaurus-A should also provide additional statistical power to confirm or refute this scenario directly.

\section{Conclusion}
\label{sec:conclusion}

With the range of viable dark matter parameter space continually shrinking and the validity of certain ``anomalies'' still under debate, it is crucial that the search for dark matter be extended to ever more exotic scenarios and/or signal sources. One such possible source may be the core of active galactic nuclei where the density of dark matter is expected to be extremely high.  Recently, several groups have explored the possibility of observing signals of dark matter from its interactions with the high-energy jets emanating from these galaxies with favorable results.  

In this work, we built upon these analyses by including the other components of the WIMP-induced gamma ray spectrum of active galactic nuclei; namely, ({\it i}) the continuum from WIMP annihilation into light standard model states which subsequently radiate and/or decay into photons and ({\it ii}) the direct (loop-induced) decay into photons.  In particular, we considered the gamma ray flux from the so-called ``chiral square'' model which is a model of two universal extra dimensions.  This model was chosen because of the interesting annihilation features it possesses (i.e., a ``forest'' of gamma ray lines).

We found that, like other models considered previously, the WIMP-induced gamma ray flux of AGN from the chiral square model exhibits several interesting features which may be observable with the Fermi gamma ray telescope.  In addtion to this, we have shown that, in conjunction with other measurements from direct and indirect detection experiments, the gamma ray spectrum from AGN can be an invaluable tool for restricting WIMP parameter space. 

\section{Acknowledgement}
M.A.G. and C.B.J. would like to thank Sangwook Park, Zdzislaw Musielak and Ohad Shemmer for useful discussions pertaining to the astrophysical aspects of this work.  G.~S. is  supported by the U. S. Department of Energy under the contract DE-FG-02-95ER40896.

\appendix

\section{Cross Section in the Collinear Approximation}
\label{app:collinear}

As is evident from Fig.~\ref{fg:xn-vs-E2}, the dominant contributions to the cross section and, thus, the gamma ray flux come from the $s$-channel Feynman diagrams due to resonances when the intermediate $E$ goes on-shell.  Thus, it appears a safe assumption would be to neglect the $u$-channel diagrams shown in Fig.~\ref{fg:feynman-diagrams} since there is no chance for the intermediate $E$ in these diagrams to go on-shell.  

Besides the resonant-enhancement from the $E$ going on-shell, there is another source of enhancement in the process considered here: a logarithmic enhancement when the photon is collinear with the final-state electron (note that, since the angle of observation does not line up with the jet axis, there is no possibility of the photon being collinear with the {\it initial-state} electron).  Below, we outline the calculation assuming that the resonant and collinear regime makes up the most important contribution to the cross section.  This calculation and this appendix follow closely Appendix B of Ref.~\cite{Profumo:2010} (by design) and provides an important check on our calculation using the full set of Feynman diagrams and the exact kinematics of the process. 

The pieces of the amplitude-squared that have the collinear log enhancement are:
\begin{eqnarray}
\frac{1}{2} \sum_{\lambda,spins} \left| {\cal M}_2 \right|^2 &=& \frac{1}{2} e^2 g_1^4 (Y_L^2 + Y_R^2)^2 \frac{E_2 E_5}{t_{45} [ (E_2 - \delta)^2 + \Gamma^2/4 ]} \left[ 1 + \frac{(Y_L^2 - Y_R^2)^2}{(Y_L^2 + Y_R^2)^2} \cos\theta \right]  \,, \\
\nonumber\\
\frac{1}{2} \sum_{\lambda,spins} 2 \left| {\cal M}_{S_2} {\cal M}_{S_3}^* \right| &=& - e^2 g_1^4 (Y_L^2 + Y_R^2)^2 \frac{(E_2 - E_5 - \delta)E_2 E_4 (E_4 + E_5)}{t_{45} [ (E_2 - \delta)^2 + \Gamma^2/4] [ (E_2 - E_5 - \delta)^2 + \Gamma^2/4]} \nonumber\\
&& \,\,\,\,\,\,\,\,\,\,\,\,\,\,\,\,\,\,\,\,\,\,\, \times \left[ 1 + \frac{(Y_L^2 - Y_R^2)^2}{(Y_L^2 + Y_R^2)^2} \cos\theta \right] \,, \\
\nonumber\\
\frac{1}{2} \sum_{\lambda,spins} 2 \left| {\cal M}_{S_1} {\cal M}_{S_2}^* \right| &=&- 2 e^2 g_1^4 (Y_L^2 + Y_R^2)^2 \frac{(E_2 - \delta)(E_2 - E_5 - \delta) E_2^2 E_4 (E_4 + E_5)}{t_{25} t_{45} [ (E_2 - \delta)^2 + \Gamma^2/4] [ (E_2 - E_5 - \delta)^2 + \Gamma^2/4]} \nonumber\\
&& \,\,\,\,\,\,\,\,\,\,\,\,\,\,\,\,\,\,\,\,\,\,\, \times \left[ 1 - \frac{4(Y_L^2 - Y_R^2)^2}{(Y_L^2 + Y_R^2)^2} \cos\theta - \frac{(Y_L^2 - Y_R^2)^2}{(Y_L^2 + Y_R^2)^2} \cos^2\theta \right] \,,
\end{eqnarray}
while the pieces of the amplitude-squared that do not have collinear log enhancements are:
\begin{eqnarray}
\frac{1}{2} \sum_{\lambda,spins} \left| {\cal M}_1 \right|^2 &=& - e^2 g_1^4 (Y_L^4 + Y_R^4) \frac{E_4 E_5}{t_{25} (E_2 - E_5 - \delta)^2 + \Gamma^2/4 } \,\\
\nonumber\\
\frac{1}{2} \sum_{\lambda,spins} \left| {\cal M}_3 \right|^2 &=& -\frac{1}{4} e^2 g_1^4 (Y_L^2 + Y_R^2)**2 \frac{E_2 E_4}{[ (E_2 - \delta)^2 + \Gamma^2/4] [ (E_2 - E_5 - \delta)^2 + \Gamma^2/4]} \,\\
\nonumber\\
\frac{1}{2} \sum_{\lambda,spins} 2 \left| {\cal M}_{S_1} {\cal M}_{S_3}^* \right| &=& e^2 g_1^4 (Y_L^2 + Y_R^2)^2 \frac{E_2 E_4 (E_2 - \delta) (E_5 - E_2)}{t_{25} t_{45} [ (E_2 - \delta)^2 + \Gamma^2/4] [ (E_2 - E_5 - \delta)^2 + \Gamma^2/4]} \nonumber\\
&& \,\,\,\,\,\,\,\,\,\,\,\,\,\,\,\,\,\,\,\,\,\,\, \times \left[ 1 + \frac{(Y_L^2 - Y_R^2)^2}{(Y_L^2 + Y_R^2)^2} \cos\theta \right] \,.
\end{eqnarray}

We group the pieces of the amplitude-squared that contain the collinear logarithm ($|{\cal M}|^2_{log}$) and those that do not ($|{\cal M}|^2_{no \, log}$) as:
\begin{eqnarray}
|{\cal M}|^2_{log} &=& \frac{1}{2} \left[ \left| {\cal M}_{S_2} \right|^2 + 2 Re | {\cal M}_{S_2}  {\cal M}_{S_3}^* | + 2 Re | {\cal M}_{S_1}  {\cal M}_{S_2}^* | \right] \, , \\
|{\cal M}|^2_{no \, log} &=& \frac{1}{2} \left[ \left| {\cal M}_{S_1} \right|^2 + \left| {\cal M}_{S_3} \right|^2 + 2 Re | {\cal M}_{S_1}  {\cal M}_{S_3}^* | \right] \, ,
\end{eqnarray}
 and compute the cross section as:
\begin{eqnarray}
\frac{d^2 \sigma}{dE_5 d\Omega_5} &=& \frac{1}{(2\pi)^5} \frac{1}{32 M_B^2 E_2} \biggl[ |{\cal M}|^2_{log} t_{45} \int d\Omega_4 \frac{E_5 E_4}{t_{45}} + 4 \pi |{\cal M}|^2_{no \, log} \biggr] \nonumber\\
&=& \frac{\pi}{(2\pi)^5} \frac{1}{32 M_B^2 E_2} \biggl[ |{\cal M}|^2_{log} t_{45} \ln \left( \frac{4 E_4^2}{m_e^2} \right) + 4 |{\cal M}|^2_{no \, log} \biggr] \,.
\end{eqnarray}

\section{Bayesian fit using Markov Chain Monte Carlo}
\label{apx:mcmc}

In this work we utilize a Markov Chain Monte Carlo (MCMC) to fit the available data. The MCMC approach is based
on Bayesian methods to scan over specified input parameters given
constraints on an output set. 
In Bayes' rule, the posterior probability of the model parameters, $\theta$, given the data, $d$, and model, $M$, is given by
\begin{equation}
p(\theta | d,M) = {\pi(\theta|M) p(d|\theta,M)\over p(d|M)},
\end{equation}
where $\pi(\theta|M)$ is known as the prior on the model parameters which contains information on the parameters before unveiling the data.  The $p(d|\theta,M)$ term is the likelihood and is given below in Eq.~\ref{eq:likelihood}.  The $p(d|M)$ term is called the evidence, but is often ignored as the probabilities are properly normalized to sum to unity.  A common algorithm for MCMCs is that of Metropolis and Hastings (MH), in which a random point, $\theta_{i}$,
is chosen in a model's parameter space and has an associated likelihood,
${\cal L}_{i}$, based on the applied constraints.  A collection of these points,
$\left\{ \theta_{i}\right\} $,
constructs the chain. The probability of choosing another point that
is different than the current one is given by the ratio of their respective
likelihoods: ${\rm min}(\frac{{\cal L}_{i+1}}{{\cal L}_{i}},1)$. Therefore,
the next proposed point is chosen if the likelihood of the next point
is higher than the current. Otherwise, the current point is repeated
in the chain.   We adopt the likelihood
\begin{equation}
{\cal L}_{i}=e^{-\Sigma_{j}\chi_{j}^{2}/2}
      =e^{-\Sigma_{j}(y_{ij}-d_{j})^{2}/2\sigma_{j}^{2}},
\label{eq:likelihood}
\end{equation}
where $y_{ij}$ are the observables calculated from the input parameters
of the $i^{th}$ chain, $d_{j}$ are the values of the experimental
and theoretical constraints and $\sigma_{j}$ are the associated uncertainties.

The advantage of a MCMC approach is that in the limit
of large chain length 
 the distribution of points, $\theta_{i}$, approaches the posterior
distribution of the modeling parameters given the constraining data.
In addition, the set formed by a function of the points in the chain,
$f(\theta_{i})$, also follows the posterior distribution of that function
of the parameters given the data.

Instead of the MH algorithm, we adopt the Goodman-Weare (GW) algorithm~\cite{MCMCGW}, which has been shown to be more efficient at choosing trial points and therefore increases speed. A collection of initially random points referred to as ``walkers'' at time $t$, $\{\theta_i(t)\}$, are placed in parameter space.  A given walker's next trial position, $\theta_i(t)^\prime$, is based chosen along a line connecting it to another randomly chosen walker, $\theta_j$, with the trial 
\be
\theta_i(t) \to \theta_i^\prime = \theta_j + Z \left( \theta_i(t) - \theta_j \right),
\ee
where $Z$ is a random variable with the distribution
\be
g(z)\propto {1\over \sqrt z}
\ee
in the interval $\left[1/a,a\right]$ and vanishes elsewhere.  The trial point is accepted if 
\be
q\equiv z^{N-1} \frac{{\cal L}_{t+1}}{{\cal L}_{t}} > r,
\ee

where $r$ is a uniform random variate.  This selection ensures detailed balance within the algorithm, which is necessary for the algorithm to be statistically relevant.  The parameter $a$ sets the scale of jumps.  The value $a=2$ is often used as it yields a very efficient exploration of the parameter space.  

We allow a ``burn-in'' period to prevent correlation of the walker's initial random state to the set of walkers used for the posterior distribution.  We guarantee this by constructing the autocorrelation time
\be
C_i(T)={1\over n-T}\sum_{t=1}^{n-T}(\theta_i(t)-\bar \theta_i)(\theta_i(t+T)-\bar \theta_i)
\ee
for each input parameter and verify that the burn-in length is at least 5 times the auto-correlation scale.


\begin{thebibliography}{99}

\bibitem{Bloom:1997vm} 
  E.~D.~Bloom and J.~D.~Wells,
  Phys.\ Rev.\ D {\bf 57}, 1299 (1998)
  [astro-ph/9706085].

\bibitem{Profumo:2010} 
  M.~Gorchtein, S.~Profumo and L.~Ubaldi,
  Phys.\ Rev.\ D {\bf 82}, 083514 (2010)
  [arXiv:1008.2230 [hep-ph]].

\bibitem{Tait:2012} 
  J.~Huang, A.~Rajamaran and T.~Tait,
  JCAP {\bf 1205}, 027 (2012)
  [arXiv:1109.2587 [hep-ph]].

\bibitem{Chang:2012sk} 
  S.~Chang, Y.~Gao and M.~Spannowsky,
  JCAP {\bf 1211}, 053 (2012)
  [arXiv:1210.1870 [astro-ph.HE]].

\bibitem{Profumo:2013jeb} 
  S.~Profumo, L.~Ubaldi and M.~Gorchtein,
  JCAP {\bf 1304}, 012 (2013)
  [arXiv:1302.1915 [astro-ph.HE]].

\bibitem{Bertone:2009cb} 
  G.~Bertone, C.~B.~Jackson, G.~Shaughnessy, T.~M.~P.~Tait and A.~Vallinotto,
  Phys.\ Rev.\ D {\bf 80}, 023512 (2009)
  [arXiv:0904.1442 [astro-ph.HE]].

\bibitem{Jackson:2009kg} 
  C.~B.~Jackson, G.~Servant, G.~Shaughnessy, T.~M.~P.~Tait and M.~Taoso,
  JCAP {\bf 1004}, 004 (2010)
  [arXiv:0912.0004 [hep-ph]].

\bibitem{Bertone:2010fn} 
  G.~Bertone, C.~B.~Jackson, G.~Shaughnessy, T.~M.~P.~Tait and A.~Vallinotto,
  JCAP {\bf 1203}, 020 (2012)
  [arXiv:1009.5107 [astro-ph.HE]].

\bibitem{Jackson:2013pjq} 
  C.~B.~Jackson, G.~Servant, G.~Shaughnessy, T.~M.~P.~Tait and M.~Taoso,
  arXiv:1302.1802 [hep-ph].

\bibitem{Jackson:2013tca} 
  C.~B.~Jackson, G.~Servant, G.~Shaughnessy, T.~M.~P.~Tait and M.~Taoso,
  arXiv:1303.4717 [hep-ph].

\bibitem{Dobrescu:2004zi} 
  B.~A.~Dobrescu and E.~Ponton,
  JHEP {\bf 0403}, 071 (2004)
  [hep-th/0401032].

\bibitem{Burdman:2005sr} 
  G.~Burdman, B.~A.~Dobrescu and E.~Ponton,
  JHEP {\bf 0602}, 033 (2006)
  [hep-ph/0506334].

\bibitem{Burdman:2006gy} 
  G.~Burdman, B.~A.~Dobrescu and E.~Ponton,
  Phys.\ Rev.\ D {\bf 74}, 075008 (2006)
  [hep-ph/0601186].

\bibitem{Dobrescu:2007xf} 
  B.~A.~Dobrescu, K.~Kong and R.~Mahbubani,
  JHEP {\bf 0707}, 006 (2007)
  [hep-ph/0703231 [HEP-PH]].

\bibitem{Dobrescu:2007ec} 
  B.~A.~Dobrescu, D.~Hooper, K.~Kong and R.~Mahbubani,
  JCAP {\bf 0710}, 012 (2007)
  [arXiv:0706.3409 [hep-ph]].

\bibitem{Gondolo:1999ef} 
  P.~Gondolo and J.~Silk,
  Phys.\ Rev.\ Lett.\  {\bf 83}, 1719 (1999)
  [astro-ph/9906391].

\bibitem{Falcone:2010fk} 
  A.~A.~Abdo {\it et al.}  [Fermi Collaboration],
  Astrophys.\ J.\  {\bf 719}, 1433 (2010)
  [arXiv:1006.5463 [astro-ph.HE]].

\bibitem{herfbh}
  C.D.~Dermer and G.~Menon
  High Energy Radiation form Black Holes: Gamma Rays, Cosmic Rays, and Neutrinos, (Princeton University Press: Princeton, NJ).



\bibitem{Belanger:2010gh} 
  G.~Belanger, F.~Boudjema, P.~Brun, A.~Pukhov, S.~Rosier-Lees, P.~Salati and A.~Semenov,
  Comput.\ Phys.\ Commun.\  {\bf 182}, 842 (2011)
  [arXiv:1004.1092 [hep-ph]].

\bibitem{Gnedin:2003rj} 
  O.~Y.~Gnedin and J.~R.~Primack,
  Phys.\ Rev.\ Lett.\  {\bf 93}, 061302 (2004)
  [astro-ph/0308385].

\bibitem{Ade:2013zuv} 
  P.~A.~R.~Ade {\it et al.}  [Planck Collaboration],
  arXiv:1303.5076 [astro-ph.CO].

\bibitem{Aprile:2012nq} 
  E.~Aprile {\it et al.}  [XENON100 Collaboration],
  Phys.\ Rev.\ Lett.\  {\bf 109}, 181301 (2012)
  [arXiv:1207.5988 [astro-ph.CO]].
    
\bibitem{fermi_line2012} 
A.~Albert ``Search for Gamma-ray Spectral Lines in the Milky Way Diffuse with the Fermi Large Area Telescope''.  The Fermi Symposium 2012.








\bibitem{Akerib:2012ak} 
  D.~S.~Akerib {\it et al.}  [LUX Collaboration],
  arXiv:1210.4569 [astro-ph.IM].

\bibitem{xenon1t} 
E.~Aprile ``XENON1T: a ton scale Dark Matter Experiment''.  UCLA Dark Matter 2010.

\bibitem{MCMCGW} 
J.~Goodman and J.~Weare,
 Comm.\ App.\ Math.\ and Comp.\ Sci.\ {\bf 5} 65 (2010)


\end{thebibliography}
\end{document}